\errorstopmode
\input amssym.def
\input amssym.tex


\magnification=\magstephalf
\hsize=14.0 true cm
\vsize=19 true cm
\hoffset=1.0 true cm
\voffset=2.0 true cm

\abovedisplayskip=12pt plus 3pt minus 3pt
\belowdisplayskip=12pt plus 3pt minus 3pt
\parindent=1.0em


\font\sixrm=cmr6
\font\eightrm=cmr8
\font\ninerm=cmr9

\font\sixi=cmmi6
\font\eighti=cmmi8
\font\ninei=cmmi9

\font\sixsy=cmsy6
\font\eightsy=cmsy8
\font\ninesy=cmsy9

\font\sixbf=cmbx6
\font\eightbf=cmbx8
\font\ninebf=cmbx9

\font\eightit=cmti8
\font\nineit=cmti9

\font\eightsl=cmsl8
\font\ninesl=cmsl9

\font\sixss=cmss8 at 8 true pt
\font\sevenss=cmss9 at 9 true pt
\font\eightss=cmss8
\font\niness=cmss9
\font\tenss=cmss10

\font\eighttt=cmtt8

\font\sixmib=cmmib6
\font\sevenmib=cmmib7
\font\eightmib=cmmib8
\font\ninemib=cmmib9
\font\tenmib=cmmib10

 at 12 true pt
 at 12 true pt
\font\bigrm=cmr10 at 12 true pt
 at 12 true pt
 at 12 true pt

 at 16 true pt
 at 16 true pt
\font\Bigrm=cmr12 at 16 true pt
 at 16 true pt
 at 16 true pt

\catcode`@=11
\newfam\ssfam
\newfam\mibfam

\def\tenpoint{\def\rm{\fam0\tenrm}%
    \textfont0=\tenrm \scriptfont0=\sevenrm \scriptscriptfont0=\fiverm
    \textfont1=\teni  \scriptfont1=\seveni  \scriptscriptfont1=\fivei
    \textfont2=\tensy \scriptfont2=\sevensy \scriptscriptfont2=\fivesy
    \textfont3=\tenex \scriptfont3=\tenex   \scriptscriptfont3=\tenex
    \textfont\itfam=\tenit                  \def\it{\fam\itfam\tenit}%
    \textfont\slfam=\tensl                  \def\sl{\fam\slfam\tensl}%
    \textfont\bffam=\tenbf \scriptfont\bffam=\sevenbf
                           \scriptscriptfont\bffam=\fivebf
                           \def\bf{\fam\bffam\tenbf}%
    \textfont\ssfam=\tenss \scriptfont\ssfam=\sevenss
                           \scriptscriptfont\ssfam=\sevenss
                           \def\ss{\fam\ssfam\tenss}%
    \textfont\mibfam=\tenmib \scriptfont\mibfam=\sevenmib
                             \scriptscriptfont\mibfam=\sevenmib
                             \def\mib{\fam\mibfam\tenmib}%
    \normalbaselineskip=13pt
    \setbox\strutbox=\hbox{\vrule height8.5pt depth3.5pt width0pt}%
    \let\big=\tenbig
    \normalbaselines\rm}

\def\ninepoint{\def\rm{\fam0\ninerm}%
    \textfont0=\ninerm      \scriptfont0=\sixrm
                            \scriptscriptfont0=\fiverm
    \textfont1=\ninei       \scriptfont1=\sixi
                            \scriptscriptfont1=\fivei
    \textfont2=\ninesy      \scriptfont2=\sixsy
                            \scriptscriptfont2=\fivesy
    \textfont3=\tenex       \scriptfont3=\tenex
                            \scriptscriptfont3=\tenex
    \textfont\itfam=\nineit \def\it{\fam\itfam\nineit}%
    \textfont\slfam=\ninesl \def\sl{\fam\slfam\ninesl}%
    \textfont\bffam=\ninebf \scriptfont\bffam=\sixbf
                            \scriptscriptfont\bffam=\fivebf
                            \def\bf{\fam\bffam\ninebf}%
    \textfont\ssfam=\niness \scriptfont\ssfam=\sixss
                            \scriptscriptfont\ssfam=\sixss
                            \def\ss{\fam\ssfam\niness}%
    \textfont\mibfam=\ninemib \scriptfont\mibfam=\sixmib
                            \scriptscriptfont\mibfam=\sixmib
                            \def\mib{\fam\mibfam\ninemib}%
    \normalbaselineskip=12pt
    \setbox\strutbox=\hbox{\vrule height8.0pt depth3.0pt width0pt}%
    \let\big=\ninebig
    \normalbaselines\rm}

\def\eightpoint{\def\rm{\fam0\eightrm}%
    \textfont0=\eightrm      \scriptfont0=\sixrm
                             \scriptscriptfont0=\fiverm
    \textfont1=\eighti       \scriptfont1=\sixi
                             \scriptscriptfont1=\fivei
    \textfont2=\eightsy      \scriptfont2=\sixsy
                             \scriptscriptfont2=\fivesy
    \textfont3=\tenex        \scriptfont3=\tenex
                             \scriptscriptfont3=\tenex
    \textfont\itfam=\eightit \def\it{\fam\itfam\eightit}%
    \textfont\slfam=\eightsl \def\sl{\fam\slfam\eightsl}%
    \textfont\bffam=\eightbf \scriptfont\bffam=\sixbf
                             \scriptscriptfont\bffam=\fivebf
                             \def\bf{\fam\bffam\eightbf}%
    \textfont\ssfam=\eightss \scriptfont\ssfam=\sixss
                             \scriptscriptfont\ssfam=\sixss
                             \def\ss{\fam\ssfam\eightss}%
    \textfont\ttfam=\eighttt \def\tt{\fam\ttfam\eighttt}%
    \textfont\mibfam=\eightmib \scriptfont\mibfam=\sixmib
                             \scriptscriptfont\mibfam=\sixmib
                             \def\mib{\fam\mibfam\eightmib}%
    \normalbaselineskip=10pt
    \setbox\strutbox=\hbox{\vrule height7.0pt depth2.0pt width0pt}%
    \let\big=\eightbig
    \normalbaselines\rm}

\def\tenbig#1{{\hbox{$\left#1\vbox to8.5pt{}\right.\n@space$}}}
\def\ninebig#1{{\hbox{$\textfont0=\tenrm\textfont2=\tensy
                       \left#1\vbox to7.25pt{}\right.\n@space$}}}
\def\eightbig#1{{\hbox{$\textfont0=\ninerm\textfont2=\ninesy
                       \left#1\vbox to6.5pt{}\right.\n@space$}}}

\font\sectionfont=cmbx10
\font\subsectionfont=cmti10

\def\figurecaptionfont{\ninepoint}
\def\tablecaptionfont{\ninepoint}
\def\footnotefont{\eightpoint}


\newcount\equationno
\newcount\bibitemno
\newcount\figureno
\newcount\tableno

\equationno=0
\bibitemno=0
\figureno=0
\tableno=0


\footline={\ifnum\pageno=0{\hfil}\else
{\hss\rm\the\pageno\hss}\fi}


\def\section #1. #2 \par
{\vskip0pt plus .10\vsize\penalty-100 \vskip0pt plus-.10\vsize
\vskip 1.6 true cm plus 0.2 true cm minus 0.2 true cm
\global\def\equationlabel{#1}
\global\equationno=0
\leftline{\sectionfont #1. #2}\par
\immediate\write\terminal{Section #1. #2}
\vskip 0.7 true cm plus 0.1 true cm minus 0.1 true cm
\noindent}


\def\subsection #1 \par
{\vskip0pt plus 1.0 true cm\penalty-50 \vskip0pt plus-1.0 true cm
\vskip2.5ex plus 0.1ex minus 0.1ex
\leftline{\subsectionfont #1}\par
\immediate\write\terminal{Subsection #1}
\vskip1.0ex plus 0.1ex minus 0.1ex
\noindent}


\def\appendix #1. #2 \par
{\vskip0pt plus .10\vsize\penalty-100 \vskip0pt plus-.10\vsize
\vskip 1.6 true cm plus 0.2 true cm minus 0.2 true cm
\global\def\equationlabel{\hbox{\rm#1}}
\global\equationno=0
\leftline{\sectionfont Appendix #1. #2}\par
\immediate\write\terminal{Appendix #1. #2}
\vskip 0.7 true cm plus 0.1 true cm minus 0.1 true cm
\noindent}



\def\equation#1{$$\displaylines{\qquad #1}$$}
\def\enum{\global\advance\equationno by 1
\hfill\llap{{\rm(\equationlabel.\the\equationno)}}}
\def\noenum{\hfill}

\def\nexteq#1{\cr\noalign{\vskip#1}\qquad}


\def\ifundefined#1{\expandafter\ifx\csname#1\endcsname\relax}

\def\ref#1{\ifundefined{#1}?\immediate\write\terminal{unknown reference
on page \the\pageno}\else\csname#1\endcsname\fi}

\newwrite\terminal
\newwrite\bibitemlist

\def\bibitem#1#2\par{\global\advance\bibitemno by 1
\immediate\write\bibitemlist{\string\def
\expandafter\string\csname#1\endcsname
{\the\bibitemno}}
\item{[\the\bibitemno]}#2\par}

\def\beginbibliography{
\vskip0pt plus .15\vsize\penalty-100 \vskip0pt plus-.15\vsize
\vskip 1.2 true cm plus 0.2 true cm minus 0.2 true cm
\leftline{\sectionfont References}\par
\immediate\write\terminal{References}
\immediate\openout\bibitemlist=biblist
\frenchspacing\parindent=1.8em
\vskip 0.5 true cm plus 0.1 true cm minus 0.1 true cm}

\def\endbibliography{
\immediate\closeout\bibitemlist
\nonfrenchspacing\parindent=1.0em}


\def\figurecaption#1{\global\advance\figureno by 1
\narrower\figurecaptionfont Fig.~\the\figureno. #1}

\def\tablecaption#1{\global\advance\tableno by 1
\centerline{\tablecaptionfont Table~\the\tableno. #1}}

\def\thicktablerule{\hrule height0.8pt}
\def\thintablerule{\hrule height0.4pt}

\tenpoint

\immediate\openin\bibitemlist=biblist
\ifeof\bibitemlist\immediate\closein\bibitemlist
\else\immediate\closein\bibitemlist
\input biblist \fi


\def\thismonth{\ifcase\month\or
January\or February\or March\or April\or May\or June\or
July\or August\or September\or October\or November\or December\fi}

\input epsf
\epsfclipon



\def\rmd{{\rm d}}

\def\rme{{\rm e}}
\def\rmO{{\rm O}}

\def\urltilde{\kern -.15em\lower .7ex\hbox{\~{}}\kern .04em}



\def\proof{\noindent{\sl Proof:}\kern0.6em}

\def\frac#1#2{\hbox{$#1\over#2$}}
\def\dual{\mathstrut^*\kern-0.1em}

\def\lvec#1{\setbox0=\hbox{$#1$}
    \setbox1=\hbox{$\scriptstyle\leftarrow$}
    #1\kern-\wd0\smash{
    \raise\ht0\hbox{$\raise1pt\hbox{$\scriptstyle\leftarrow$}$}}
    \kern-\wd1\kern\wd0}
\def\rvec#1{\setbox0=\hbox{$#1$}
    \setbox1=\hbox{$\scriptstyle\rightarrow$}
    #1\kern-\wd0\smash{
    \raise\ht0\hbox{$\raise1pt\hbox{$\scriptstyle\rightarrow$}$}}
    \kern-\wd1\kern\wd0}
\def\cvec#1{\kern-0.5pt\vec{\kern0.5pt #1}}

\def\slash#1{\setbox2=\hbox{$\displaystyle#1$}%
             \setbox3=\hbox{$\displaystyle/$}%
             #1\kern-0.8\wd2/\kern-1.0\wd3\kern0.8\wd2\kern0.5pt}

\def\wick#1{\setbox2=\hbox{$\displaystyle#1$}
    \setbox3=\null\ht3=3.0pt\dp3=0.0pt\wd3=20.0pt
    #1\kern-\wd2\kern3.0pt\raise11.0pt\vbox{\hrule height0.3pt
    \hbox{\vrule width0.3pt\box3\vrule width0.3pt}}\kern-24.0pt\kern\wd2}

\def\longwick#1{\setbox2=\hbox{$\displaystyle#1$}
    \setbox3=\null\ht3=3.0pt\dp3=0.0pt\wd3=27.0pt
    #1\kern-\wd2\kern3.0pt\raise11.0pt\vbox{\hrule height0.3pt
    \hbox{\vrule width0.3pt\box3\vrule width0.3pt}}\kern-31.0pt\kern\wd2}

\def\verylongwick#1{\setbox2=\hbox{$\displaystyle#1$}
    \setbox3=\null\ht3=3.0pt\dp3=0.0pt\wd3=43.0pt
    #1\kern-\wd2\kern3.0pt\raise11.0pt\vbox{\hrule height0.3pt
    \hbox{\vrule width0.3pt\box3\vrule width0.3pt}}\kern-47.0pt\kern\wd2}


\def\nab#1{{\nabla_{#1}}}
\def\nabstar#1{{\nabla\kern0.5pt\smash{\raise 4.5pt\hbox{$\ast$}}
               \kern-5.5pt_{#1}}}

\def\drvstar#1{{\partial\kern0.5pt\smash{\raise 4.5pt\hbox{$\ast$}}
               \kern-6.0pt_{#1}}}
\def\sdrvstar#1{{\partial\kern0.4pt\smash{\raise 3.6pt\hbox{$\ast$}}
                \kern-4.8pt_{#1}}}

\def\ldrvstar#1{{\lvec{\,\partial}\kern-0.5pt\smash{\raise 4.5pt\hbox{$\ast$}}
               \kern-5.0pt_{#1}}}


\def\MSbar{\overline{\rm MS\kern-0.5pt}\kern0.5pt}




\def\diracstar#1#2{
    \setbox0=\hbox{$\gamma$}\setbox1=\hbox{$\gamma_{#1}$}
    \gamma_{#1}\kern-\wd1\kern\wd0
    \smash{\raise4.5pt\hbox{$\scriptstyle#2$}}}


\def\SUthree{{\rm SU(3)}}
\def\SUn{{\rm SU}(N)}

\def\sun{\frak{su}(N)}
\def\tr{{\rm tr}}


\def\SG{S_{\rm G}}

\def\Spf{S_{\rm pf}}

\def\ct{{\rm ct}}


\def\Dw{D_{\rm w}}
\def\Dwdag{{\Dw}\kern-4pt^{\dagger}\kern1pt}
\def\Dm{D}
\def\Dmdag{\Dm^{\dagger}\kern-1pt}


\def\eps{\epsilon}

\def\om{\omega}
\def\obs{{\cal O}}
\def\qb{\hat{q}}
\def\pb{\hat{p}}
\def\pib{\hat{\pi}}
\def\Ub{\hat{U}}
\def\Vb{\hat{V}}
\def\omb{\hat{\om}}
\def\nub{\hat{\nu}}
\def\sigb{\hat{\sigma}}
\def\norm#1{\left\|#1\right\|}

\def\Hs{{\cal H}}
\def\mbnd{\kappa}
\def\Op{\Delta}
\def\Oph{\hat{\Op}}
\def\gbar{\bar{g}}
\def\kfact{k}
\def\ams{\alpha_s}

\def\ct{c_{\rm t}}
\def\var{{\rm var}}

%
\rightline{CERN-TH-2017-057}
\vskip1.2cm
\centerline{\Bigrm SMD-based numerical stochastic perturbation theory}

\vskip 0.6 true cm
\centerline{\bigrm Mattia Dalla Brida$^{\rm a}$ and
Martin L\"uscher$^{\rm b,c}$}

\vskip1.5ex
\centerline{{\it $^{\rm a}$Dipartimento di Fisica,
Universit\`a di Milano-Bicocca and}}
\centerline{{\it INFN, Sezione di Milano-Bicocca,
Piazza della Scienza 3, I-20126 Milano, Italy}}

\vskip1.0ex
\centerline{{\it $^{\rm b}$CERN,
Theoretical Physics Department, 1211 Geneva 23, Switzerland}}

\vskip1.0ex
\centerline{{\it $^{\rm c}$Albert Einstein Center for Fundamental Physics}}
\centerline{{\it Institute for Theoretical Physics,
Sidlerstrasse 5, 3012 Bern, Switzerland}}

\vskip 0.8 true cm
\thintablerule
\vskip 2.0ex
\ninepoint
\leftline{\bf Abstract}
\vskip 1.0ex\noindent
The viability of a variant of numerical stochastic perturbation theory,
where the Langevin equation is replaced by the SMD algorithm, is examined.
In particular, the convergence of the process
to a unique stationary state is rigorously established
and the use of higher-order symplectic integration schemes is shown
to be highly profitable in this context.
For illustration, the gradient-flow coupling in finite volume
with Schr\"odinger functional boundary conditions
is computed to two-loop (i.e.~NNL) order in the SU(3) gauge theory.
The scaling behaviour of the algorithm turns out to be rather favourable
in this case, which allows the computations to be driven close to the continuum
limit.
\vskip 2.0ex
\thintablerule

\tenpoint


\section 1. Introduction

Numerical stochastic perturbation theory (NSPT)
[\ref{SPThI}--\ref{DiRenzoScorzato}] is
a powerful tool that
allows many interesting calculations in QCD
and other quantum field theories to be performed
to high order in the interactions.
For technical reasons,
the computations proceed in the framework of lattice field theory,
but results for renormalized quantities in
the continuum theory can then be obtained
through an extrapolation to vanishing lattice spacing.
NSPT can be highly automated and the application of the method
in finite volume and to correlation
functions of complicated composite fields
gives rise to hardly any additional difficulties.

Reliable extrapolations to the continuum limit
require accurate data at several lattice spacings
in the scaling region.
NSPT calculations can therefore rapidly become large-scale
projects, where computational efficiency is all-important.
Traditionally, NSPT is based on the Langevin equation, but the success of
the HMC algorithm [\ref{HMC}]
in lattice QCD suggests that
the inclusion of a molecular-dynamics update step in the underlying
stochastic process might be beneficial.
Smaller autocorrelation times and an improved
scaling behaviour towards the continuum limit
could perhaps be achieved in this way.
Moreover, through the use of highly efficient
symplectic integration schemes,
the systematic errors deriving from the discretization of the
simulation time may conceivably be reduced.

NSPT based on the SMD (stochastic
molecular dynamics, or generalized HMC) algorithm [\ref{Horowitz}]
has recently been briefly looked at in ref.~[\ref{LatKobe}]
and was found to perform well.
Here we establish
the convergence of the algorithm to a unique
stationary state and study its efficiency in
the case of the gradient-flow coupling in the SU(3) gauge theory.
Various technical problems are addressed
along the way, among them the modifications required to ensure
that the stochastic process does not run away in the gauge directions.

\section 2. Stochastic molecular dynamics

In order to bring out the basic structure of
the SMD-variant of NSPT most clearly,
a generic system described by a set $q=(q_1,\ldots,q_n)$ of
real coordinates and an action $S(q)$ is considered
in this and the following two sections.

\subsection 2.1 Preliminaries

The action $S(q)$ is assumed to be differentiable
and to have an expansion
in powers of a coupling $g$ of the form
\equation{
  S(q)=\sum_{r=0}^{\infty} g^rS_r(q),
  \enum
}
where $S_r(q)$ is a polynomial in $q$ of degree $d_r\geq2$.
Moreover, it is taken for granted that the leading-order term
\equation{
  S_0(q)=\frac{1}{2}(q,\Op q)=\frac{1}{2}\sum_{k,l=1}^nq_k\Op_{kl}q_l
  \enum
}
is a strictly positive quadratic form in the coordinates.

The observables $\obs(q)$ of interest
are assumed to be similarly expandable
and their expectation values
\equation{
  \langle\obs\rangle
  ={1\over{\cal Z}_S}\int \rmd q_1\ldots\rmd q_n\,\obs(q)\kern0.5pt\rme^{-S(q)}
  \enum
}
then have a well-defined perturbation expansion with coefficients
given by Feynman diagrams as usual.

\subsection 2.2 SMD algorithm

The SMD algorithm
operates in the phase space of the theory and
thus updates both the coordinates $q$ and their canonical momenta
$p=(p_1,\ldots,p_n)$.
An SMD update cycle consists of a momentum rotation
followed by a molecular-dynamics evolution and, optionally,
an acceptance-rejection step.

The momenta are
rotated in a random direction according to
\equation{
  p\to c_1p+c_2\upsilon,
  \enum
}
where the momentum $\upsilon$ is randomly chosen
from a Gaussian distribution with mean zero and unit
variance. The coefficients
\equation{
  c_1=\rme^{-\gamma\eps},
  \qquad
  c_2=(1-c_1^2)^{1/2},
  \enum
}
depend on the simulation time step $\eps>0$ and a parameter
$\gamma>0$ that controls the rotation angle.

In the second step, the molecular-dynamics equations
\equation{
  \partial_tp=-\nabla S(q),
  \qquad
  \partial_tq=p,
  \enum
}
are integrated from the current simulation time $t$ to $t+\eps$
using a reversible symplectic integration scheme (see subsect.~2.3).
The algorithm (momentum rotation followed by the molecular-dynamics
evolution) simulates the canonical distribution
\equation{
  {1\over{\cal Z}_H}\rme^{-H(p,q)},\qquad
  H(p,q)=\frac{1}{2}(p,p)+S(q),
  \enum
}
provided the integration errors are negligible or
an acceptance-rejection step is included in the update cycle
which corrects for these [\ref{JansenLiu}].
Stochastic estimates of the expectation values (2.3)
of the observables of interest are then obtained by
averaging their values over a range of simulation time.

\subsection 2.3 Integration schemes

The molecular-dynamics equations may be integrated by
applying a sequence of the elementary steps
\equation{
  I_{p,h}:\;p\to p-h\nabla S(q),
  \enum
  \nexteq{2.0ex}
  I_{q,h}:\;q\to q+hp,
  \enum
}
to the current momenta and coordinates, with
step sizes $h$ proportional to $\eps$. A well-known example
is the leapfrog integrator
$I_{p,\eps/2}I_{q,\eps}I_{p,\eps/2}$,
and several highly efficient schemes are described in
ref.~[\ref{OMF}].

Integrators $I_{\eps}$ of this kind are symplectic and they can be
(and are here) required to be reversible, i.e.~to be such that
\equation{
  I_{\eps}PI_{\eps}=P,
  \enum
}
where $P$ stands for the momentum reflection $p\to-p$.

\section 3. Stochastic perturbation theory

Stochastic perturbation theory [\ref{SQI},\ref{SQII}]
is usually derived from the Langevin
equation by expanding the stochastic variables
and the driving forces in powers of the coupling.
In this section,
another (although probably closely related)
form of stochastic perturbation theory is discussed,
which is obtained by expanding the SMD algorithm in the same way.

\subsection 3.1 SMD algorithm at weak coupling

Since the acceptance-rejection step is not smooth in the coupling,
its effects would be difficult to take into account in
perturbation theory.
In the following, the acceptance-rejection step is therefore omitted,
without further notice,
and one is thus left with an algorithm
that simulates the system only up to integration errors.

The histories $p(t),q(t)$ of the momenta and coordinates generated
by the SMD algorithm depend on the coupling $g$ through the force term in
the integration step (2.8). In particular, they
are smooth functions of the coupling and
may consequently be expanded in the asymptotic series
\equation{
  p(t)=\sum_{r=0}^{\infty}g^r\pb_r(t),
  \qquad
  q(t)=\sum_{r=0}^{\infty}g^r\qb_r(t),
  \enum
}
where the leading-order histories $\pb_0(t),\qb_0(t)$ coincide with
the ones generated by the algorithm in the free
theory with action (2.2).

In terms of the coefficients $\pb_r,\qb_r$, the momentum rotation (2.4)
becomes
\equation{
  \pb_r\to c_1\pb_r+\delta_{r0}c_2\upsilon
  \enum
}
and the molecular-dynamics integration steps (2.8),(2.9)
assume the form
\equation{
  I_{\pb_r,h}:\;\pb_r\to\pb_r-h\hat{F}_r(\qb_0,\ldots,\qb_r),
  \enum
  \nexteq{2.5ex}
  I_{\qb_r,h}:\;\qb_r\to\qb_r+h\pb_r.
  \enum
}
The forces $\hat{F}_r$ in eq.~(3.3) are given by
\equation{
  \nabla S(q)=\sum_{r=0}^{\infty}
  g^r\hat{F}_r(\qb_0,\ldots,\qb_r)
  \enum
}
and it is understood that
all momenta and all coordinates are updated alternately
so that the variables
on the right of eqs.~(3.3),(3.4) are always the current ones.
For any given initial data, these rules completely determine
the histories $\pb_r(t),\qb_r(t)$.

\subsection 3.2 Perturbation expansion of observables

Similarly to the gradient of the action, any observable
\equation{
  \obs(q)=\sum_{r=0}^{\infty}
  g^r\hat{\obs}_r(\qb_0,\ldots,\qb_r)
  \enum
}
may be expanded in powers of the coupling.
The coefficients $k_r(\obs)$ in the perturbation expansion
\equation{
  \langle\obs\rangle=\sum_{r=0}^{\infty}k_r(\obs)g^r
  \enum
}
of its expectation value (2.3) then coincide
with the averages of $\hat{\obs}_r(\qb_0(t),\ldots,\qb_r(t))$
over the simulation time $t$
up to statistical (and integration) errors.

\section 4. Convergence to a stationary state

Stochastic processes can run away or do not converge
to a stationary distribution for other reasons.
In the case of the stochastic perturbation
theory described in sect.~3, the asymptotic stationarity
of the underlying process can be rigorously shown if the
simulation step size $\eps$ is sufficiently small.
The range of step sizes, where convergence is
guaranteed, depends on the chosen integration scheme for
the molecular-dynamics equations and the matrix $\Op$ in
the leading-order part (2.2) of the action.

\subsection 4.1 Molecular-dynamics evolution in the free theory

If the coupling $g$ is turned off, the molecular-dynamics equations
become linear and their (approximate) integration
from time $t$ to $t+\eps$ amounts
to a linear transformation
\equation{
  \pmatrix{p\cr q\cr}\to M\pmatrix{p\cr q\cr}
  \enum
}
of the current momenta and coordinates.
The $2n\times 2n$ matrix $M$ in this equation has a block structure,
\equation{
  M=\pmatrix{M_{pp}& M_{pq}\cr
             M_{qp}& M_{qq}\cr},
  \enum
}
with $n\times n$ blocks that are polynomials in $\eps$ and
the matrix $\Op$ with some numerical coefficients.
In particular, they are commuting real symmetric matrices.
In the case of the leapfrog integrator,
\equation{
  M_{pp}=M_{qq}=1-\frac{1}{2}\eps^2\Op,
  \quad
  M_{pq}=-\eps\Op\bigl(1-\frac{1}{4}\eps^2\Op\bigr),
  \quad
  M_{qp}=\eps,
 \enum
}
and explicit expressions for the blocks can be obtained
for other popular integrators as well (see appendix A).

The symplecticity and reversibility of the chosen integration scheme
imply
\equation{
  M_{pp}=M_{qq},
  \enum
  \nexteq{2.0ex}
  M_{pp}M_{qq}-M_{pq}M_{qp}=1.
  \enum
}
It follows from these relations that the Hamilton function
\equation{
  \hat{H}(p,q)=\frac{1}{2}(p,p)+\frac{1}{2}(q,\Oph q),
  \enum
  \nexteq{2.5ex}
  \Oph=-M_{pq}(M_{qp})^{-1}=\Op+\rmO(\eps),
  \enum
}
is exactly conserved by the integrator.
In the following, $\eps$ and $\Op$ are assumed to be such that
$M_{qp}$ is non-singular and $\Oph$ positive definite.
Both conditions are met in the case of the leapfrog integrator
if $\eps^2\norm{\Op}<4$, where
$\norm{\Op}$ is the largest eigenvalue of $\Op$.
The probability distribution
\equation{
  \hat{P}(p,q)\propto\rme^{-\hat{H}(p,q)}
  \enum
}
is then well-behaved and preserved
by the SMD algorithm, since it
is preserved both by the momentum rotation
and the molecular-dynamics evolution.

\subsection 4.2 Convergence of the leading-order process

For any
initial distribution of the momenta and coordinates,
the SMD algorithm produces a sequence of distributions, which
converges to the stationary distribution (4.8) in the free theory.
One can show this by working out the
action of an SMD update cycle on a given distribution, but
the convergence of the algorithm may be established
more easily starting from the identity
\equation{
  \pmatrix{p(t)\cr q(t)}=\tilde{M}^{t/\eps}\pmatrix{p(0)\cr q(0)}
  +c_2\sum_{u=0}^{t-\eps}\tilde{M}^{(t-u)/\eps-1}M\pmatrix{\upsilon(u)\cr 0},
  \enum
}
where $\upsilon(u)$ is the momentum chosen randomly in the
momentum rotation (2.4) at simulation time $u=0,\eps,2\eps,\ldots$ and
\equation{
  \tilde{M}=M\pmatrix{c_1 & 0\cr 0 & 1\cr}.
  \enum
}
The long-time behaviour of the momenta and coordinates thus depends
on the properties of the matrix $\tilde{M}$.

The blocks $M_{pp},M_{pq},M_{qp}$ and $\Oph$ are commuting real
symmetric matrices.
Since
\equation{
  M_{pp}^2=1-M_{qp}\Oph M_{qp}
  \enum
}
and since $\Oph$ is assumed to be positive definite, the eigenvalues
of $M_{pp}$ have magnitude strictly less than $1$.
The eigenvalues of $\tilde{M}$ are then
\equation{
  \lambda_{\pm}=\frac{1}{2}\bigl\{
  (1+c_1)\mu\pm\sqrt{(1+c_1)^2\mu^2-4c_1}
  \bigr\},
  \enum
}
where $\mu$ runs through the eigenvalues of $M_{pp}$.
In particular, $|\lambda_{\pm}|<1$ and $\tilde{M}$ is thus
a contraction matrix.

The first term on the right of eq.~(4.9)
consequently dies away exponentially with increasing
simulation time $t$.
Since the random momenta are normally distributed,
the momenta and coordinates at large $t$ are then normally distributed
as well, with mean zero and variances equal to their two-point autocorrelation
functions.
In the large time limit, the $qq$ autocorrelation function, for example,
is given by
\equation{
  \langle q(t)q(s)\rangle_{\upsilon}
  \mathrel{\mathop=_{t\geq s}}
  \bigl\{\tilde{M}^{(t-s)/\eps}K\bigr\}_{qq},
  \enum
  \nexteq{2.5ex}
  K=c_2^2
  \sum_{u=0}^{\infty}
  \tilde{M}^{u/\eps}MP_{+}
  \bigl\{\tilde{M}^{u/\eps}M\bigr\}^T,
  \qquad
  P_{+}=\pmatrix{1&0\cr0&0\cr},
  \enum
}
and the other two-point functions by the $pp$, $pq$ and $qp$
blocks of the matrix on the right of eq.~(4.13).
The kernel $K$ satisfies
\equation{
  \tilde{M}K\tilde{M}^T=K-c_2^2MP_{+}M^T,
  \enum
}
i.e.~an inhomogeneous linear equation that has a unique solution,
since $\tilde{M}$ is a con\-traction matrix.
A few lines of algebra then show that
\equation{
  K=\pmatrix{1&0\cr0&\Oph^{-1}}
  \enum
}
solves the equation.
In particular, the variances of the momenta and coordinates at
any fixed simulation time coincide with the ones of the stationary
distribution (4.8), which proves that the latter coincides
with the distribution simulated by the SMD algorithm.

\subsection 4.3 Convergence beyond the leading order

The assumed structure of the action $S(q)$ implies that the
force in the molecular-dynamics integration step (3.3) is of the form
\equation{
  \hat{F}_r(\qb_0,\ldots,\qb_r)=
  \Op\qb_r+\hat{F}'_r(\qb_0,\ldots,\qb_{r-1}),
  \enum
}
where the second term on the right is a polynomial in the
coordinates up to order $r-1$. If the history of the latter and
the associated momenta is already known (including their values at
the intermediate integration steps), the
integration of the molecular-dynamics equations at order $r$
thus amounts to solving an inhomogeneous linear recursion.
A moment of thought then reveals that
\equation{
  \pmatrix{\pb_r(t)\cr\qb_r(t)}=
  \tilde{M}^{t/\eps}\pmatrix{\pb_r(0)\cr\qb_r(0)}
  +\sum_{u=\eps}^t
  \tilde{M}^{(t-u)/\eps}
  \hat{V}_r(\pb_0(u),\ldots,\qb_{r-1}(u))
  \enum
}
for all $r\geq1$. The ``vertices''
$\hat{V}_r(\pb_0,\qb_0;\ldots;\pb_{r-1},\qb_{r-1})$ in this formula
are polynomials in their arguments, whose exact form depends
on both the integration scheme and the forces (4.17).

Recalling eq.~(4.9) and the fact that $\tilde{M}$ is a contraction
matrix,
the convergence of the autocorrelation functions of the
momenta and coordinates at large times $t$ may now be shown
recursively from order $0$ to any finite order $r$.
Equation (4.18) actually
allows the highest-order variables in any correlation function
to be expressed through lower-order ones up to an exponentially
decaying contribution. Clearly, in the large-time limit,
the autocorrelation functions do not depend on the initial
distribution of the variables and are stationary, i.e.~invariant
under time translations.

The expectation values of
the coefficients in the perturbation expansion
(3.6) of the observables coincide with a sum of autocorrelation
functions of the coordinates at equal times.
Their convergence at large times is therefore guaranteed as well.

\subsection 4.4 Summary

The discussion in this section shows that the
SMD algorithm converges to all orders of perturbation theory if
the matrix $\Op$ is strictly positive and if $\eps^2\norm{\Op}<\mbnd$,
where $\mbnd$ depends on the molecular-dynamics integrator.
In the case of the leapfrog, the 2nd-order OMF and the
4th-order OMF integrators,
$\mbnd$ is equal to $4$, $6.51$ and $9.87$, respectively
(cf.~appendix A).

\section 5. Stochastic perturbation theory in lattice QCD

With respect to the generic system considered so far,
the situation in lattice QCD is complicated by the gauge symmetry
and the quark fields. In this section, stochastic perturbation
theory is first set up for the pure $\SUn$ gauge theory.
The modifications required for the damping of the gauge
modes are then discussed and the section ends with
a brief description of how the quarks can be included in the simulations.

\subsection 5.1 Lattice fields

The lattice theory is set up
on a $T\times L^3$ lattice
with periodic boundary conditions in the space directions and
Schr\"odinger functional (SF)
[\ref{SF},\ref{SFquark}] or open-SF [\ref{openSF}]
boundary conditions in the time direction.
In both cases the link variables $U(x,\mu)$ satisfy
\equation{
  U(x,k)=1,\quad k=1,2,3,
  \enum
}
at time $x_0=T$ and additionally at $x_0=0$ if SF boundary
conditions are chosen. The notation used in the following
coincides with the one previously employed in ref.~[\ref{openSF}].
In particular, all dimensionful quantities are expressed in
units of the lattice spacing.

The momenta $\pi(x,\mu)$ of the link variables $U(x,\mu)$ take values
in the Lie algebra $\sun$
of the gauge group. They vanish at the boundaries of the lattice,
where the link variables are frozen to unity.
The scalar product of any two momentum fields,
\equation{
  (\pi,\upsilon)=
  \sum_{x,\mu}w_{x,\mu}^{-1}\pi^a(x,\mu)\upsilon^a(x,\mu),
  \enum
}
includes a conventional weight factor
\equation{
  w_{x,\mu}=\cases{2 & if $\mu>0$ and $x_0=0$ or $x_0=T$,\cr
                   \noalign{\vskip0.5ex}
                   1 & otherwise,\cr}
  \enum
}
which will reappear below in various expressions.

Infinitesimal gauge transformations are fields
$\om(x)$ of $\sun$ elements defined on the sites $x$ of the lattice.
They vanish at $x_0=T$ and must, furthermore,
be constant at $x_0=0$
if SF boundary conditions are imposed.
In this latter case, the fields may be split in two parts according to
\equation{
  \om(x)=\left(1-x_0/T\right)\om(0)+\hat{\om}(x),
  \enum
}
where $\hat{\om}(x)$ vanishes at $x_0=0,T$ and is otherwise unconstrained.
A possible choice of scalar product is then
\equation{
  (\om,\nu)=TL^3\om^a(0)\nu^a(0)+\sum_x\hat{\om}^a(x)\hat{\nu}^a(x),
  \enum
}
while in the case of open-SF boundary conditions
\equation{
  (\om,\nu)=\sum_xw_x^{-1}\om^a(x)\nu^a(x)
  \enum
}
with $w_x=2$ at $x_0=0,T$ and $w_x=1$ elsewhere.

Independently of the boundary conditions imposed on the gauge field,
the quark fields are required to vanish at both $x_0=0$ and $x_0=T$.
No weight factor is included in the scalar product of these fields.

\subsection 5.2 Basic stochastic process

In the absence of quark fields,
the SMD algorithm proceeds along the lines of sect.~2.
For the action $S(U)$ of the gauge field any of the
frequently used ones may be taken. The Hamilton function
from which the SMD algorithm derives is then given by
\equation{
  H(\pi,U)=\frac{1}{2}(\pi,\pi)+S(U).
  \enum
}
In particular, the random field in the momentum rotation
\equation{
  \pi(x,\mu)\to c_1\pi(x,\mu)+c_2\upsilon(x,\mu)
  \enum
}
must be distributed with probability density proportional to
$\rme^{-{1\over2}(\upsilon,\upsilon)}$.

If the weight factor $w_{x,\mu}$ in the scalar product (5.2)
is also included in the symplectic structure (i.e.~the Poisson bracket),
the molecular-dynamics equations assume the form
\equation{
  \partial_t\pi(x,\mu)=-g_0w_{x,\mu}(\partial^a_{x,\mu}S)(U)T^a,
  \qquad
  \partial_tU(x,\mu)=g_0\pi(x,\mu)U(x,\mu).
  \enum
}
For later convenience, the expressions on the right of the equations
have been scaled by the bare gauge coupling $g_0$, an operation that
could be undone by rescaling the simulation time.
The integration steps (2.8) and (2.9) are then given by
\equation{
  I_{\pi,h}:\;\pi(x,\mu)\to \pi(x,\mu)
  -hg_0w_{x,\mu}(\partial^a_{x,\mu}S)(U)T^a,
  \enum
  \nexteq{2.0ex}
  I_{U,h}:\;U(x,\mu)\to\rme^{hg_0\pi(x,\mu)}U(x,\mu).
  \enum
}
Since it will be omitted in perturbation theory,
the acceptance-rejection step needed to correct
for the integration errors is not described here.

\subsection 5.3 Perturbation expansion

In perturbation theory, the link variables are represented by
a gauge potential $A_{\mu}(x)$ through
\equation{
  U(x,\mu)=\exp\{g_0A_{\mu}(x)\}=
  1+\sum_{r=0}^{\infty}g_0^{r+1}\Ub_r(x,\mu),
  \qquad
  \Ub_0(x,\mu)=A_{\mu}(x).
  \enum
}
The gauge potential takes values in $\sun$ and satisfies the same
boundary conditions as the momentum field
\equation{
  \pi(x,\mu)=\sum_{r=0}^{\infty}g_0^r\pib_r(x,\mu).
  \enum
}
When the gauge and momentum fields are replaced by these expansions,
the SMD algorithm leads to a hierarchy of stochastic equations
as in the case of the generic system considered in sect.~3.

At lowest order in the coupling,
the momentum is rotated according to eq.~(5.8)
(with $\pib_0$ in place of $\pi$)
and the integration steps (5.10),(5.11) become
\equation{
  I_{\pib_0,h}:\;\pib_0(x,\mu)\to\pib_0(x,\mu)-h(\Delta\Ub_0)(x,\mu),
  \enum
  \nexteq{2.0ex}
  I_{\Ub_0,h}:\;\Ub_0(x,\mu)\to\Ub_0(x,\mu)+h\pib_0(x,\mu),
  \enum
}
where $\Delta$ is the symmetric linear operator in the leading-order
expression
\equation{
  S_0(U)=\frac{1}{2}(A,\Delta A)
  \enum
}
for the gauge action.

\subsection 5.4 Damping of the gauge modes

The space $\Hs_1$ of gauge potentials may be split into the subspace
$\Hs_1^L$ of gauge modes and its orthogonal complement $\Hs_1^T$.
There is a one-to-one correspondence between the
infinitesimal gauge transformations, introduced in subsect.~5.1,
and the gauge modes through the forward difference operator
\equation{
  (d\nu)(x,\mu)=\partial_{\mu}\nu(x),
  \enum
}
which maps any field $\nu$ in the space $\Hs_0$
of infinitesimal gauge transformations to a field $d\nu\in\Hs_1^L$.
The adjoint operator $d^{\ast}$ goes in the opposite direction and
is defined by the requirement that
\equation{
  (d\nu,A)=(\nu,d^{\ast}\!A)
  \enum
}
for all $\nu\in\Hs_0$ and $A\in\Hs_1$. In particular, the subspace
$\Hs_1^T$ coincides with the space of
gauge potentials $A$ satisfying $d^{\ast}\!A=0$
($d^{\ast}$ is given explicitly in appendix B).

Since the gauge modes are annihilated by $\Delta$,
the $\Hs_1^L$-component of the
leading-order gauge field $\Ub_0$ performs a random walk
and thus slowly runs away
in the course of a simulation.
Stability can be regained by applying a gauge transformation
\equation{
  \pi(x,\mu)\to \Lambda(x)\pi(x,\mu)\Lambda(x)^{-1},
  \qquad \Lambda(x)=\exp\{\eps g_0\om(x)\},
  \enum
  \nexteq{2.5ex}
  U(x,\mu)\to\Lambda(x)U(x,\mu)\Lambda(x+\hat{\mu})^{-1},
  \enum
}
to the fields after each SMD update cycle, where
$\om\in\Hs_0$ is a new field that is evolved together with the other
fields. There are two update steps for this field, the first,
\equation{
  \om(x)\to c_1\om(x),
  \enum
}
being applied together with the momentum rotation and
the second,
\equation{
  \om(x)\to\om(x)+\eps\lambda_0(d^{\ast}\kern-0.5pt C)(x),
  \enum
  \nexteq{2.5ex}
  C_{\mu}(x)={1\over2g_0}\Bigl\{
  U(x,\mu)-U(x,\mu)^{-1}-{1\over N}\tr\bigl[U(x,\mu)-U(x,\mu)^{-1}\bigr]
  \Bigr\},
  \enum
}
at the end of the molecular-dynamics evolution.
The parameter $\lambda_0>0$ controls the feedback from the
current gauge field to the gauge-damping field $\om(x)$
and can, in principle,
be set to any value. Clearly, the history of
gauge-invariant observables is not affected by
these modifications of the SMD algorithm,
but as discussed below (in subsect.~5.5),
they have the desired damping effect on the gauge modes%
\kern1pt\footnote{$\dagger$}{\footnotefont%
In the continuous-time limit $\eps\to0$, the time-dependence
of the gauge-damping field is governed by a first-order differential
equation and the modified algorithm integrates a form of
the stochastic molecular-dynamics equations,
which coincides with the standard one up to a
time-dependent gauge transformation (see appendix C).}.

Like the other fields, the gauge-damping field
is expanded in a series
\equation{
  \om(x)=\sum_{r=0}^{\infty}g_0^r\omb_r(x)
  \enum
}
in perturbation theory. At leading order, the new update steps
are then
\equation{
  \omb_0(x)\to c_1\omb_0(x),
  \enum
  \nexteq{2.0ex}
  \omb_0(x)\to\omb_0(x)+\eps\lambda_0(d^{\ast}\kern-0.5pt\Ub_0)(x),
  \enum
  \nexteq{2.0ex}
  \Ub_0(x,\mu)\to\Ub_0(x,\mu)-\eps(d\omb_0)(x,\mu),
  \enum
}
and the higher-order rules have the familiar hierarchical
structure.

\subsection 5.5 Long-time stationarity of the process

Although the stochastic process now includes variables without
associated momenta, the discussion in sect.~4 applies here too
with only minor changes.
In particular,
the convergence of the process to a unique stationary state
is guaranteed to all orders of the coupling, if
the matrix $\tilde{M}$ describing the evolution of the fields
$(\pib_0,\Ub_0,\omb_0)$
at leading order is a contraction matrix.

Since the gauge modes are zero modes of $\Delta$,
the subspace of field vectors
\equation{
  (\pib_0,\Ub_0,\omb_0)=
  (d\nub_0,d\sigb_0,\omb_0),
  \qquad
  \nub_0,\sigb_0\in\Hs_0,
  \enum
}
is invariant under the action of $\tilde{M}$.
When restricted to its orthogonal complement (which is
invariant too),
the matrix describes the
evolution of the $\Hs_1^T$-components of $\pib_0,\Ub_0$.
The chosen boundary conditions (SF or open-SF) imply
the strict positivity of $\Delta$ in this subspace
and $\tilde{M}$ is therefore a contraction matrix there
if
\equation{
  \eps^2\|\Delta\|<\mbnd,
  \enum
}
where $\mbnd$ depends on the molecular-dynamics integrator (cf.~subsect.~4.4).

In the subspace (5.28) of the gauge modes, on the other hand,
the action of $\tilde{M}$
amounts to applying the linear transformation
\equation{
  \pmatrix{\nub_0\cr\sigb_0\cr\omb_0}
  \to
  \pmatrix{c_1 & 0 & 0 \cr
           \eps c_1 & 1-\eps^2\lambda_0d^{\ast}\kern-1.0pt d & -\eps c_1 \cr
           0 & \eps\lambda_0d^{\ast}\kern-1.0pt d & c_1}
  \pmatrix{\nub_0\cr\sigb_0\cr\omb_0}
  \enum
}
to the fields.
The associated eigenvalues are equal to $c_1$ or to
\equation{
  \frac{1}{2}\bigl\{
  b\pm\sqrt{b^2-4c_1}\bigr\},
  \qquad
  b=1+c_1-\eps^2\lambda_0\mu^2,
  \enum
}
where $\mu^2$ is any eigenvalue of $d^{\ast}\kern-1.0pt d$.
This operator has no zero modes and
some simple estimates then show that $\tilde{M}$ is
a contraction matrix in the subspace of gauge modes, provided the
bound
\equation{
  \eps^2\lambda_0\|d^{\ast}\kern-1.0pt d\|<2(1+c_1)
  \enum
}
holds.

Convergence of the stochastic process to equilibrium is thus guaranteed
if the inequalities (5.29) and (5.32) are both
satisfied. With the chosen boundary conditions,
\equation{
  \|\Delta\|\leq 16k,
  \qquad
  \|d^{\ast}\kern-1.0pt d\|\leq16,
  \enum
}
where $k$ is equal to $1,5/3$ and $3.648$ in the case of the Wilson
[\ref{Wilson}],
tree-level Symanzik improved [\ref{SymImpI},\ref{OnShell}]
and Iwasaki [\ref{Iwasaki}] gauge action, respectively.
In practice, the constraints (5.29) and (5.32) therefore tend
to be fairly mild and the main concern is to ensure that
the integration errors are sufficiently small at the
chosen values of $\eps$.

\subsection 5.6 Inclusion of the quark fields

As in the case of the HMC algorithm [\ref{HMC}], the effects of
the quarks can be included in SMD simulations by adding a
multiplet of
pseudo-fermion fields to the theory with the appropriate action.
The details are not important here and it suffices
to know that the action
is a sum of terms, one per pseudo-fermion field $\phi$,
of the form
\equation{
  \Spf(U,\phi)=(R(U)\phi,R(U)\phi),
  \enum
}
with $R(U)$ being some gauge-covariant linear operator
(see ref.~[\ref{LesHouches}], for example).

Only few modifications of the SMD update cycle are required
in the presence of the pseudo-fermion fields. Similarly to the
momentum field, each field is
rotated according to\kern1pt\footnote{$\dagger$}{\footnotefont%
There is no reason other than simplicity to set
the parameter $\gamma$ that determines the rotation angle
to the same value for all fields.}
\equation{
  \phi(x)\to c_1\phi(x)+c_2(R(U)^{-1}\eta)(x)
  \enum
}
at the beginning of the cycle, where $\eta$ is a Gaussian random field
with mean zero and unit variance.
The molecular-dynamics evolution of the gauge and momentum field
then proceeds at fixed pseudo-fermion fields, with the contribution
of the pseudo-fermion action to the driving force
properly taken into account. Clearly, the gauge transformation
(5.19),(5.20) applied at the end of the update cycle must
be extended to the pseudo-fermion fields.

When the algorithm is expanded in powers of the coupling $g_0$, the
renormalization of the quark masses should be taken into account
so that the masses in the leading-order
stochastic equations are the renormalized ones.
The pseudo-fermion fields de\-couple from the other fields
at lowest order
and are simulated exactly by the random rotation (5.35).
Convergence of the stochastic process to a unique stationary
state is then again
guaranteed to all orders,
if the bounds (5.29) and (5.32) are satisfied.

\section 6. Computation of the gradient-flow coupling

The gradient-flow coupling in finite volume with SF boundary conditions
has recently been used in step scaling studies of three-flavour QCD
[\ref{StepQCDI},\ref{StepQCDII}]. Such calculations
serve to relate the low-energy properties of the theory to the
high-energy regime, where contact with the standard
QCD parameters and matrix elements
can be made in perturbation theory [\ref{StepScaling}]
(see ref.~[\ref{HouchesII}] for a review).

In the following, the perturbation expansion of the
gradient-flow coupling is determined in the $\SUthree$
Yang--Mills theory up to two-loop order,
using the SMD-variant of NSPT described in the previous
section. To one-loop order, the expansion coefficient
in the $\MSbar$ scheme of
dimensional regularization is known in infinite volume
since a while [\ref{WilsonFlow}], but
a huge effort plus the
best currently available techniques were required to
be able to extend this calculation to the next order
[\ref{HarlanderNeumann}].
In finite volume with SF boundary conditions,
these techniques do not apply and a similar analytical computation
may be practically infeasible at present.

\subsection 6.1 Definition of the coupling

The renormalized coupling considered in this paper belongs to a
family of couplings based on the Yang--Mills gradient flow.
Explicitly, it is given by [\ref{FritzschRamos}]
\equation{
  \gbar^2=\kfact\left\{t^2\langle E(t,x)\rangle
  \right\}_{T=L,x_0=L/2,\sqrt{8t}=cL},
  \enum
}
where $E(t,x)$ denotes the Yang--Mills action density at flow time $t$
and position $x$,
$c$ is a parameter of the scheme and the proportionality constant
$\kfact$ is determined
by the requirement that $\gbar^2$ coincides with $g_0^2$ to
lowest order of perturbation theory. Most of the time $c$ will be
set to $0.3$, which implies a localization range of the action
density of about $0.3\times L$.

On the lattice, the gradient flow is implemented as in ref.~[\ref{openSF}].
The Wilson action,
with boundary terms so as to ensure the absence of O($a$) lattice
effects in the flow equation, is thus used to generate the flow.
For the action density $E(t,x)$ in eq.~(6.1) either
the Wilson action density or the square of the familiar symmetric ``clover''
expression for the gauge-field tensor is inserted.
Furthermore, alternative couplings, where the full action density
is replaced by its spatial or time-like part, are considered as well.

All in all this makes 6 different action densities and couplings,
labeled  w, ws, wt, c, cs, and ct,
where the letters stand for
Wilson, clover, space and time, respectively
($E^{\rm ct}$, for example,
denotes the clover expression for the time-like part of
the action density).

\subsection 6.2 Expansion in powers of $\alpha_s$

The gradient flow coupling
has an expansion
\equation{
  \gbar^2=4\pi\left\{\ams+k_1\ams^2+k_2\ams^3+\ldots\right\}
  \enum
}
in powers of the running coupling $\ams$
in the $\MSbar$ scheme of dimensional regularization
at momentum scale $q$,
with coefficients $k_1,k_2,\ldots$ depending on $q$ and $L$.
If $q$ is scaled with $L$ like
\equation{
  q=(cL)^{-1},
  \enum
}
the coefficients are of order $1$ and independent of $L$
in the continuum theory [\ref{RenFlow}].

In the following $k_1$ and $k_2$ are computed by combining the expansion
\equation{
  t^2\langle E(t,x)\rangle
  =E_0g_0^2+E_1g_0^4+E_2g_0^6+\ldots
  \enum
}
obtained in NSPT with the relation
\equation{
  \alpha_s=\alpha_0+d_1\alpha_0^2+d_2\alpha_0^3+\ldots,
  \quad
  \alpha_0={g_0^2\over4\pi},
  \enum
}
between $\ams$ and the bare coupling.
For the Wilson gauge action (which is the action
used in NSPT),
the coefficients $d_1$ and $d_2$
are accurately known [\ref{LatMSbar}].
The coefficients $k_1,k_2$ determined in this way
depend on the spacing of the simulated lattices so that
an extrapolation to the continuum limit is then still required.

\subsection 6.3 Computation of the coefficients $E_k$ in NSPT

In order to cancel the O($a$) lattice effects in $E_k$,
the action of the theory must include boundary counterterms at
$x_0=0$ and $x_0=T$ with a tuned coupling
[\ref{SF}]\kern1pt\footnote{$\dagger$}{\footnotefont%
In ref.~[\ref{openSF}] the improvement coefficient $\ct$ is denoted
by $c_{\rm G}'$.}
\equation{
  \ct=1+\ct^{(1)}g_0^2+\ct^{(2)}g_0^4+\ldots.
  \enum
}
In the case of the Wilson action,
\equation{
  \ct^{(1)}=-0.08900(5),
  \qquad
  \ct^{(2)}=-0.0294(3),
  \enum
}
as was shown long ago [\ref{CtI}--\ref{CtIII}]. The
counterterms lead to further terms in the forces
that drive the molecular-dynamics evolution, but
do not require a modification of the general framework described
in sect.~5. Alternatively, the effects of the counterterms
can be included in the calculations by treating $\ct-1$ as a second
coupling.

Once a representative sample of gauge configurations (5.12) has been
generated, the stochastic estimation of the
gradient-flow coupling requires $E(t,x)$ to be expanded in
powers of $g_0$ for each of these configurations.
To this end,
the gauge field $V_t(x,\mu)$ at flow time $t$ is represented by a gauge
potential $B_{\mu}(t,x)$ according to
\equation{
  V_t(x,\mu)=\exp\{g_0B_{\mu}(t,x)\}
  =1+\sum_{r=0}^{\infty}g_0^{r+1}\Vb_{t,r}(x,\mu).
  \enum
}
The numerical integration of the flow equation, using
a Runge--Kutta integrator, for example,
then amounts to applying a sequence of integration steps
to the expansion coefficients of the field
as in the case of the integration of the
molecular-dynamics equations. Gauge damping
is however not required here, since the linearized flow is
transversal and leaves the gauge modes unchanged.

\topinsert
\newdimen\digitwidth
\setbox0=\hbox{\rm 0}
\digitwidth=\wd0
\catcode`@=\active
\def@{\kern\digitwidth}
\tablecaption{Simulation parameters}
\vskip-1.0ex

$$\vbox{\settabs\+&%
                  xxx&xi&%
                  xxxxxxxxx&xx&%
                  xxxxxxxxx&xx&%
                  xxxxxxxxx&xx&%
                  xxxxxxxxx&xx\cr
\thicktablerule
\vskip1.2ex
                \+& \hfill $L$\hfill
                 && \hfill $\gamma$\hfill
                 && \hfill $\eps$\hfill
                 && \hfill $\Delta t_{\rm ms}/\eps$\hfill
                 && \hfill $N_{\rm ms}$\hfill
                 &\cr
\vskip0.8ex
\thintablerule
\vskip1.2ex
{\ninepoint
  \+& \hfill $10$\hfill
  &&  \hfill $5.0$\hfill
  &&  \hfill $0.168$\hfill
  &&  \hfill $190$\hfill
  &&  \hfill $59400$\hfill
  &\cr
  \+& \hfill $12$\hfill
  &&  \hfill $5.0$\hfill
  &&  \hfill $0.168$\hfill
  &&  \hfill $190$\hfill
  &&  \hfill $59880$\hfill
  &\cr
  \+& \hfill $14$\hfill
  &&  \hfill $5.0$\hfill
  &&  \hfill $0.168$\hfill
  &&  \hfill $190$\hfill
  &&  \hfill $59400$\hfill
  &\cr
  \+& \hfill $16$\hfill
  &&  \hfill $5.0$\hfill
  &&  \hfill $0.168$\hfill
  &&  \hfill $190$\hfill
  &&  \hfill $59880$\hfill
  &\cr
  \+& \hfill $18$\hfill
  &&  \hfill $4.5$\hfill
  &&  \hfill $0.168$\hfill
  &&  \hfill $240$\hfill
  &&  \hfill $58800$\hfill
  &\cr
  \+& \hfill $20$\hfill
  &&  \hfill $4.0$\hfill
  &&  \hfill $0.168$\hfill
  &&  \hfill $290$\hfill
  &&  \hfill $59400$\hfill
  &\cr
  \+& \hfill $24$\hfill
  &&  \hfill $3.0$\hfill
  &&  \hfill $0.190$\hfill
  &&  \hfill $270$\hfill
  &&  \hfill $71880$\hfill
  &\cr
  \+& \hfill $28$\hfill
  &&  \hfill $3.0$\hfill
  &&  \hfill $0.144$\hfill
  &&  \hfill $340$\hfill
  &&  \hfill $79200$\hfill
  &\cr
  \+& \hfill $32$\hfill
  &&  \hfill $3.0$\hfill
  &&  \hfill $0.126$\hfill
  &&  \hfill $480$\hfill
  &&  \hfill $88400$\hfill
  &\cr
  \+& \hfill $40$\hfill
  &&  \hfill $3.0$\hfill
  &&  \hfill $0.100$\hfill
  &&  \hfill $950$\hfill
  &&  \hfill $80100$\hfill
  &\cr
}
\vskip1.2ex
\thicktablerule
}
$$
\vskip0.0ex
\endinsert

\subsection 6.4 Simulation parameters and tables of results

The parameters of the NSPT runs performed for the ``measurement''
of $E_0,E_1,E_2$ are listed in table~1. In all cases, the
gauge-damping parameter $\lambda_0$ was set to $2$ and the
4th-order OMF integrator
was employed for the molecular-dynamics equations (cf.~appendix A).
Measurements were made after every $\Delta t_{\rm ms}/\eps$
update cycles using a 3rd-order Runge--Kutta integrator for
the gradient flow [\ref{WilsonFlow}], with step size varying from
$0.002$ at small flow times to $0.1$ at large times.
With this scheme, the gradient-flow
integration errors are guaranteed to be
completely negligible with respect to the statistical errors.
The number $N_{\rm ms}$ of measurements
made is listed in the last column of table~1
(the programs that were used in these simulations can be downloaded from
{\tt http://luscher.web.cern.ch/luscher/NSPT}).

At the chosen values of the parameters,
the bounds (5.29) and (5.32) are satisfied by a wide margin
so that the convergence of the SMD algorithm is rigorously guaranteed.
The results obtained on the simulated lattices
for the expansion coefficients $k_1$ and $k_2$ and their statistical errors
are listed in appendix D, for $c\in\{0.2,0.3,0.4\}$ and all choices
w,c,ws,$\ldots$ of the action density (cf.~subsect.~6.1).

\section 7. Statistical and systematic errors

The values of $k_1$ and $k_2$
obtained in NSPT depend on the scheme parameter $c$,
the lattice size $L$ (in lattice units),
the simulation step size $\eps$ and the SMD parameter $\gamma$.
No attempt is made here to determine all these dependencies in
detail. Instead some basic facts and empirical results are discussed
that helped controlling the errors in the case of the simulations
reported in this paper (see refs.~[\ref{LatKobe},\ref{PhiFour}]
for related com\-plementary studies of the $\phi^4$ theory).

\subsection 7.1 Autocorrelations and statistical variances

Usually the variances of the observables are a property of the
simulated field theory and hence independent of the simulation
algorithm. In NSPT this is not so, because the algorithm is
not exact, but mainly because the square of the order-$r$
coefficient $\hat{\obs}_r$
of an observable $\obs$ in general does not coincide with the order-$2r$
coefficient of another observable.
The time average of $\hat{\obs}_r^2$ and thus the variance
of $\hat{\obs}_r$ are then
not necessarily determined by the theory alone.
For illustration, the dependence on $\gamma$ of the
variances of the coefficients in
\equation{
  {t^2\over L^3}\sum_{\mib x}E(t,x)
  =\hat{E}_0g_0^2+\hat{E}_1g_0^4+\hat{E}_2g_0^6+\ldots
  \enum
}
(where $x_0=L/2$ and $\sqrt{8t}=cL$ as before) is shown
in fig.~1.
As will be discussed in subsect.~7.3, the integration errors
are negligible with respect to
the rapid growth of the variances
of the one- and two-loop coefficients seen at small $\gamma$,
which is therefore entirely an effect of the change of algorithm.

\topinsert
\vbox{
\vskip0ex

\epsfxsize=8.0cm\hskip2.0cm\epsfbox{plots/var1.eps}

\vskip2.5ex
\figurecaption{%
Variances of $\hat{E}^{\rm c}_k$ at fixed $L=16$, $\ct=1$, $c=0.3$
and $\eps=0.238$ versus $\gamma$.
The dotted lines connecting the data points are drawn to guide the eye.
Beyond $\gamma=3$
the variances are practically constant.
}
}
\vskip0ex
\endinsert

In practice one would like to minimize
the computational work required to obtain the
calculated coefficients to a given statistical precision.
The simulation algorithm should thus be tuned so as to minimize
the products
\equation{
  \tau_{\rm int}(\hat{\obs}_r)\times\var(\hat{\obs}_r)
  \enum
}
of the integrated autocorrelation times and variances of the
order-$r$ coefficients $\hat{\obs}_r$ of the observables
$\obs$ of interest. Empirical studies show
that the two factors often
work against each other, i.e.~algorithms tuned for small
autocorrelations tend to give large variances and vice versa.

The autocorrelation times of the coefficients $\hat{E}_k^{\rm c}$,
for example, grow monotonically with $\gamma$ (see table~2).
At the chosen point in parameter space,
the associated products (7.2)
are then minimized at values of $\gamma$ around
$0.5$, $1.0$ and $2.0$ for $k=0$, $1$ and $2$.
The example shows that there may be
no uniformly best choice of $\gamma$,
but a range of values, where all
coefficients of interest are obtained reasonably efficiently.

\topinsert
\newdimen\digitwidth
\setbox0=\hbox{\rm 0}
\digitwidth=\wd0
\catcode`@=\active
\def@{\kern\digitwidth}
\tablecaption{Autocorrelation times
of $\hat{E}_k^{\rm c}$
and associated products (7.2)\kern1pt$\dagger$}
\vskip-3.5ex

$$\vbox{\settabs\+&%
                  xxxx&xxx&%
                  xxxxx&&%
                  xxxxxxxxxx&xxx&%
                  xxxxx&&%
                  xxxxxxxxxx&xxx&%
                  xxxxx&&%
                  xxxxxxxxxx&xx\cr
\thicktablerule
\vskip1.2ex
                \+& \hfill $\gamma$\hfill
                 &&
                 && $k=0$\hfill
                 &&
                 && $k=1$\hfill
                 &&
                 && $k=2$\hfill
                 &\cr
\vskip0.8ex
\thintablerule
\vskip1.2ex
{\ninepoint
  \+& \hfill $0.5$\hfill
  &&  \hfill $@2.0$\hfill
  &&  \hfill $1.6\times10^{-6}$\hfill
  &&  \hfill $@2.0$\hfill
  &&  \hfill $9.4\times10^{-6}$\hfill
  &&  \hfill $@2.0$\hfill
  &&  \hfill $8.8\times10^{-5}$\hfill
  &\cr
  \+& \hfill $1.0$\hfill
  &&  \hfill $@2.7$\hfill
  &&  \hfill $2.2\times10^{-6}$\hfill
  &&  \hfill $@3.0$\hfill
  &&  \hfill $5.2\times10^{-6}$\hfill
  &&  \hfill $@2.9$\hfill
  &&  \hfill $1.0\times10^{-5}$\hfill
  &\cr
  \+& \hfill $2.0$\hfill
  &&  \hfill $@5.9$\hfill
  &&  \hfill $4.9\times10^{-6}$\hfill
  &&  \hfill $@6.7$\hfill
  &&  \hfill $6.8\times10^{-6}$\hfill
  &&  \hfill $@7.3$\hfill
  &&  \hfill $8.1\times10^{-6}$\hfill
  &\cr
  \+& \hfill $2.5$\hfill
  &&  \hfill $@6.7$\hfill
  &&  \hfill $5.6\times10^{-6}$\hfill
  &&  \hfill $@8.4$\hfill
  &&  \hfill $8.1\times10^{-6}$\hfill
  &&  \hfill $@9.2$\hfill
  &&  \hfill $9.1\times10^{-6}$\hfill
  &\cr
  \+& \hfill $3.0$\hfill
  &&  \hfill $@7.7$\hfill
  &&  \hfill $6.2\times10^{-6}$\hfill
  &&  \hfill $@9.4$\hfill
  &&  \hfill $8.4\times10^{-6}$\hfill
  &&  \hfill $11.2$\hfill
  &&  \hfill $9.8\times10^{-6}$\hfill
  &\cr
  \+& \hfill $5.0$\hfill
  &&  \hfill $10.9$\hfill
  &&  \hfill $8.7\times10^{-6}$\hfill
  &&  \hfill $13.0$\hfill
  &&  \hfill $1.1\times10^{-5}$\hfill
  &&  \hfill $16.6$\hfill
  &&  \hfill $1.3\times10^{-5}$\hfill
  &\cr
  \+& \hfill $9.0$\hfill
  &&  \hfill $14.8$\hfill
  &&  \hfill $1.2\times10^{-5}$\hfill
  &&  \hfill $17.7$\hfill
  &&  \hfill $1.4\times10^{-5}$\hfill
  &&  \hfill $21.5$\hfill
  &&  \hfill $1.5\times10^{-5}$\hfill
  &\cr
}
\vskip1.2ex
\thicktablerule

\vskip1.2ex
\+\hskip2pt$\dagger$
{\footnotefont%
All lattice and algorithm parameters are as in fig.~1.
The autocorrelation times}
&\cr
\vskip-0.5ex
\+\hskip1.03em{\footnotefont%
are given in units of simulation time.}
&\cr
}
$$
\vskip-2.0ex
\endinsert

\subsection 7.2 Critical slowing down

The behaviour of the autocorrelation times
and variances near the continuum limit $L\to\infty$ depends on
the simulation algorithm and the observables considered.
When NSPT is based on the Langevin equation, the autocorrelations
of the coefficients of multiplicatively renormalizable quantities
can be shown to diverge proportionally to $L^2$
[\ref{Zinn},\ref{ZinnZwanziger}], while
their variances grow at most logarithmically [\ref{LangNotes}].

At large $\gamma$, the variant of NSPT studied here
is expected to behave similarly,
since the SMD algorithm then effectively integrates the Langevin
equation.
Choosing $\gamma$ to depend on $L$ in some particular way
may, however, conceivably lead to an improved scaling behaviour.
In the free theory, for example, the autocorrelation times
grow proportionally to $L$ rather than $L^2$ if
$\gamma$ is scaled like $1/L$ [\ref{Horowitz}].
Beyond the leading order, the situation is complicated by the
algorithm-dependence of the variances and the effects of the
parameter tuning are then not easy to predict.

Considering the fact that the computational cost
of the measurements of $\hat{E}_k$ tends to be larger
than the one of the SMD update cycles,
the parameters of the runs on the large
lattices (those with $L=24,\ldots,40$ in table~1) were chosen
so that subsequent measurements are practically
uncorrelated.
At fixed $\gamma=3$ the required increase with $L$
of the measurement time separation
$\Delta t_{\rm ms}$ then turned out to be quite moderate.
Moreover, from $L=24$ to $L=40$,
the variances of $\hat{E}_k$
grow only slowly
(at $c=0.3$ and for $k=0$, $1$ and $2$
by about $2$, $19$ and $30$ percent).

\subsection 7.3 Integration errors

As already noted in appendix A, the theory is very accurately
simulated at leading order if the 4th-order OMF integrator is used
for the molecular-dynamics equations.
The expectation values of the coefficient $\hat{E}_0$ calculated
in the runs listed in table~1 in fact all
agree with the known analytic formula [\ref{FritzschRamos},\ref{openSF}]
within a relative statistical precision of about $2\times10^{-4}$.

Beyond the leading order, the integration errors remain difficult to
detect in empirical studies (see fig.~2).
The stability bounds (5.29),(5.32) are respected in all
these runs and the coefficients $\hat{H}_0,\ldots,\hat{H}_4$ of the
Hamilton function $H$ are accurately conserved,
with deficits decreasing like $\eps^5$
(as has to be the case in the asymptotic regime of
a 4th-order integrator).
It thus seems safe
to conclude that the integration errors in the tests
reported in fig.~2 are smaller than the statistical errors.

Apart from adjustments to match the target statistics,
the step sizes in the runs listed in table~1 were, for $L\geq24$,
scaled proportionally to $1/L$ so
that the integration errors fall off like $1/L^4$
at large $L$ and thus much more rapidly
than the leading lattice effects.
Since the statistical errors are approximately the same
in all runs, the scaling of the step sizes may be a luxury,
but was applied here as a safeguard measure against
an enhancement of systematic errors through the continuum extrapolation.

\topinsert
\vbox{
\vskip0ex

\epsfxsize=9.5cm\hskip1.25cm\epsfbox{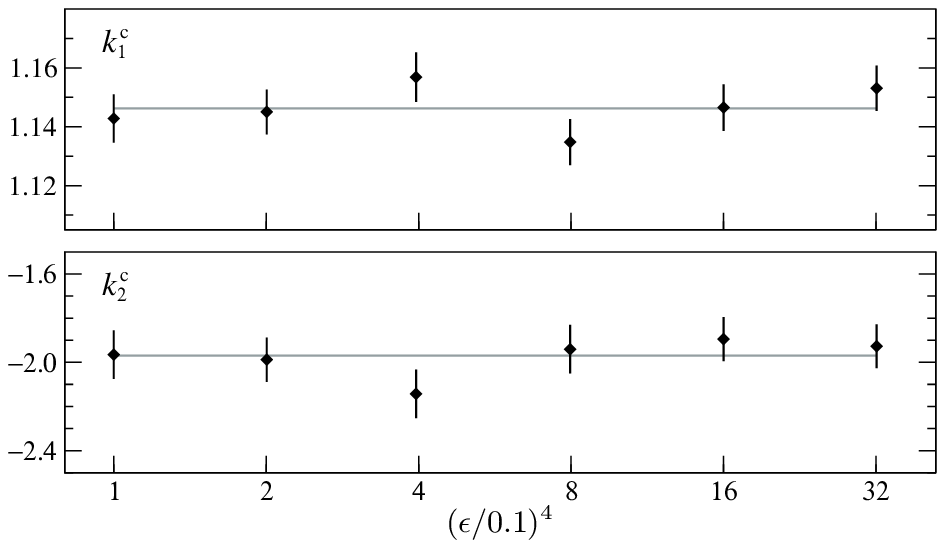}

\vskip2.0ex
\figurecaption{%
Dependence of $k^{\rm c}_1$ and $k^{\rm c}_2$ at $L=24$,
$\ct=1$ and $c=0.3$ on the simulation step size $\eps$
(note the log scale on the abscissa). In the range of data shown,
$\eps$ increases from $0.1$ to $0.238$. Each data point
is based on approximately $10^4$ statistically independent measurements
of the coefficients
$\hat{E}^{\rm c}_0$, $\hat{E}^{\rm c}_1$ and $\hat{E}^{\rm c}_2$.
}
}
\vskip0ex
\endinsert

\subsection 7.4 Extrapolation to the continuum limit

With $\rmO(a)$-improvement in place,
the $L$-dependence of the one-loop coefficient $k_1^{\rm w}$
is asymptotically given by
[\ref{SymI},\ref{SymII}]
\equation{
  k^{\rm w}_1\mathrel{\mathop=_{L\to\infty}}
  a_0+\{a_1+b_1\ln L\}L^{-2}+\rmO(L^{-3}),
  \enum
}
where the leading-order term, $a_0$, coincides with the coefficient
$k_1$ in the continuum theory. The available data are well described
by this asymptotic expression down to $L=10$ if $c\geq0.3$ and $L=12$
if $c=0.2$ (see fig.~3).
Note the fact that the data points in fig.~3 are uncorrelated and that
random fluctuations by more than one standard deviation
are consequently not improbable in a sample of
$10$ points. Plots of the other one-loop coefficients
$k_1^{\rm c},k_1^{\rm ws},\ldots$ look much the same.

\topinsert
\vbox{
\vskip0ex

\epsfxsize=10.0cm\hskip1.0cm\epsfbox{plots/c0.3-k1w.eps}

\vskip2.0ex
\figurecaption{%
Extrapolation of the simulation results for $k_1^{\rm w}$ at $c=0.3$
to the continuum limit (open point).
The full line is obtained by fitting the data in the range $10\leq L\leq40$
with the asymptotic expression (7.3).
A linear fit in the range $16\leq L\leq40$
(dashed grey line) lies practically on top of the full line.
}
}
\vskip0.0ex
\endinsert

Even though only a short extrapolation is required to reach the continuum
limit, there is no way the extrapolation errors can be rigorously assessed.
Following standard practice, the errors are estimated by
varying the fit procedure within reasonable bounds. In particular,
fits are discarded if the quality of the fit is low or
if the fit parameters are poorly determined.
Another indication of the size of the extrapolation errors
is provided by the deviations of the results obtained using
the ``w'' and the ``c'' data, respectively.
The errors of the results quoted in table~3 include both the
statistical and an estimate of the extrapolation errors.

\topinsert
\vbox{
\vskip0ex

\epsfxsize=10.0cm\hskip1.0cm\epsfbox{plots/c0.3-k2w.eps}

\vskip1.5ex
\figurecaption{%
Extrapolation of the simulation results for $k_2^{\rm w}$
at $c=0.3$ to the continuum limit (open point).
The full line is obtained by fitting the data in the range $10\leq L\leq40$
according to eq.~(7.4), with the two lowest modes of the
design matrix projected away. In the range $16\leq L\leq40$ the data can also be
fitted by a straight line (dashed grey line).
}
}
\vskip1.0ex
\endinsert

With increasing loop order,
the continuum extrapolations tend to become more difficult,
partly because one loses statistical precision and partly
because the asymptotic formulae describing the leading lattice
effects have more and more terms.
In particular, with respect to the one-loop coefficients,
the coefficients at the next order are obtained
about $10$ times less precisely and their asymptotic form
\equation{
  k^{\rm w}_2=a_0+\{a_1+b_1\ln L+c_1(\ln L)^2\}L^{-2}+\rmO(L^{-3})
  \enum
}
includes an additional logarithm.
Combinations of the logarithms in this expression only too
easily accurately approximate a constant in the range
of the available data,
but may blow up closer to the continuum limit and then strongly influence
the result of the extrapolation.

In the present context, the goal is not to determine the
values of the coefficients $a_1$, $b_1$ and $c_1$, but to
perform a short extrapolation of the data to the continuum limit.
A sensible way to stabilize fits of the data by the asymptotic
expression (7.4) is then to constrain the minimization
of the likelihood function to the directions in parameter space orthogonal
to its flattest directions (see fig.~4). The results quoted in
table~3 were obtained in this way and by varying the fit procedure,
as in the case of the one-loop coefficients, in order to assess the
extrapolation errors.

\subsection 7.5 Miscellaneous remarks

\noindent
{\it Lattice effects.}
Since the smoothing range of the gradient flow decreases
with $c$, it is no surprise that
the coefficients $k_1,k_2$
calculated in NSPT are found to be increasingly
sensitive to lattice effects when $c$ is lowered.
The continuum extrapolation of the data
then becomes more and more difficult and eventually requires
larger lattices to be simulated.

\topinsert
\newdimen\digitwidth
\setbox0=\hbox{\rm 0}
\digitwidth=\wd0
\catcode`@=\active
\def@{\kern\digitwidth}
\tablecaption{Values of $k_1$ and $k_2$ in the continuum limit}
\vskip-4.0ex

$$\vbox{\settabs\+&%
                  xxxx&xi&%
                  xxxxxxxxx&xi&%
                  xxxxxxxxx&xi&%
                  xxxxxxxxx&xi&%
                  xxxxxxxxx&xi&%
                  xxxxxxxxx&xi&%
                  xxxxxxxxx&\cr
\thicktablerule
\vskip1.2ex
                \+& \hfill $c$\hfill
                 && \hfill $k_1$\hfill
                 && \hfill $k_1^{\rm s}$\hfill
                 && \hfill $k_1^{\rm t}$\hfill
                 && \hfill $k_2$\hfill
                 && \hfill $k_2^{\rm s}$\hfill
                 && \hfill $k_2^{\rm t}$\hfill
                 &\cr
\vskip0.8ex
\thintablerule
\vskip1.2ex
{\ninepoint
  \+& \hfill $0.2$\hfill
  &&  \hfill $1.089(6)$\hfill
  &&  \hfill $1.106(7)$\hfill
  &&  \hfill $1.066(7)$\hfill
  &&  \hfill $-1.21(6)$\hfill
  &&  \hfill $-1.29(6)$\hfill
  &&  \hfill $-1.12(7)$\hfill
  &\cr
  \+& \hfill $0.3$\hfill
  &&  \hfill $1.112(5)$\hfill
  &&  \hfill $1.220(6)$\hfill
  &&  \hfill $1.005(7)$\hfill
  &&  \hfill $-1.76(4)$\hfill
  &&  \hfill $-2.17(5)$\hfill
  &&  \hfill $-1.36(6)$\hfill
  &\cr
  \+& \hfill $0.4$\hfill
  &&  \hfill $1.297(5)$\hfill
  &&  \hfill $1.685(6)$\hfill
  &&  \hfill $0.935(6)$\hfill
  &&  \hfill $-3.06(6)$\hfill
  &&  \hfill $-4.78(7)$\hfill
  &&  \hfill $-1.47(8) $\hfill
  &\cr
}
\vskip1.2ex
\thicktablerule
}
$$
\vskip0.0ex
\endinsert

\vskip1ex
\noindent
{\it Infinite-volume limit.}
The gradient-flow coupling in infinite volume runs with the flow
time $t$ and may be expanded in powers of $\ams$ at scale $q=(8t)^{-1/2}$,
as in eq.~(6.2),
the one- and two-loop coefficients in the continuum limit being
$k_1=1.0978(1)$ [\ref{WilsonFlow}] and $k_2=-0.9822(1)$
[\ref{HarlanderNeumann}].
In the finite-volume scheme considered in this paper,
and after passing to the continuum limit,
the infinite-volume limit is reached by sending $c$ to zero.
The results listed in table~3 cannot be reliably
extrapolated to $c=0$, but
the values of $k_1,\ldots,k^{\rm t}_2$ at $c=0.2$ are actually
already quite close to the infinite-volume values quoted above.

\vskip1ex
\noindent
{\it Three-loop $\beta$-function.}
The $L$-dependence of the gradient-flow
coupling $\alpha=\gbar^2/4\pi$
is governed by the renormalization group equation
\equation{
  L{\partial \alpha\over\partial L}=
  \beta_0\alpha^2+\beta_1\alpha^3+\beta_2\alpha^4+\ldots,
  \enum
}
where $\beta_0=11(2\pi)^{-1}$ and $\beta_1=51(2\pi)^{-2}$ are
the universal one- and two-loop coefficients of the
Callan--Symanzik $\beta$-function. Using the results quoted
in table~3, the three-loop coefficient may be calculated
in a few lines (see table~4). The coefficient thus has opposite
sign and is significantly larger than in the $\MSbar$ scheme,
particularly so at $c=0.4$ and if the spatial part of
the Yang--Mills action density is inserted in the definition
(6.1) of the coupling.

\section 8. Conclusions

In stochastic perturbation theory the fields are represented by
a series of coefficient fields that solve the underlying stochastic
equation order by order in the interactions. Beyond the leading order,
the probability distributions of the coefficient
fields are however not a priori known
and actually depend on the chosen stochastic process.
The variances of the observables of interest must
consequently be expected to vary with the parameters
of the simulation algorithm, an effect that tends to
considerably complicate the situation with respect to the one in
simulations based on importance sampling.

\topinsert
\newdimen\digitwidth
\setbox0=\hbox{\rm 0}
\digitwidth=\wd0
\catcode`@=\active
\def@{\kern\digitwidth}
\tablecaption{Ratio $\rho_2=\beta_2/\beta_0$ of coefficients
of the $\beta$-function}
\vskip-1.0ex

$$\vbox{\settabs\+&%
                  xxxx&xi&%
                  xxxxxxxxx&xx&%
                  xxxxxxxxx&xx&%
                  xxxxxxxxx&\cr
\thicktablerule
\vskip1.2ex
                \+& \hfill $c$\hfill
                 && \hfill $\rho_2$\hfill
                 && \hfill $\rho_2^{\rm s}$\hfill
                 && \hfill $\rho_2^{\rm t}$\hfill
                 &\cr
\vskip0.8ex
\thintablerule
\vskip1.2ex
{\ninepoint
  \+& \hfill $0.2$\hfill
  &&  \hfill $-2.38(6)$\hfill
  &&  \hfill $-2.51(6)$\hfill
  &&  \hfill $-2.22(7)$\hfill
  &\cr
  \+& \hfill $0.3$\hfill
  &&  \hfill $-2.99(4)$\hfill
  &&  \hfill $-3.74(5)$\hfill
  &&  \hfill $-2.29(6)$\hfill
  &\cr
  \+& \hfill $0.4$\hfill
  &&  \hfill $-4.88(6)$\hfill
  &&  \hfill $-8.04(7)$\hfill
  &&  \hfill $-2.21(8)$\hfill
  &\cr
}
\vskip1.2ex
\thicktablerule
}
$$
\vskip0.0ex
\endinsert

The SMD algorithm provides a technically attractive basis for NSPT.
Compared to the traditional version of
NSPT [\ref{SPThI}--\ref{DiRenzoScorzato}],
where the starting point
is the Langevin equation, a significantly
improved efficiency is achieved,
particularly so near the continuum limit.
Moreover, the available
highly accurate integrators for the molecular-dynamics equations
allow the integration errors to be easily kept under control.
For the reasons mentioned above,
some tuning of the friction parameter $\gamma$ is however required
and must take into account the variances of the
observables of interest.

The inclusion of the quark fields in the SMD algorithm along the lines of
sect.~5 is straightforward and is not expected to slow down the simulations
by a large factor [\ref{DiRenzoScorzato}]. In general, the cost
of NSPT computations very much depends on the observables considered,
the order in perturbation theory and the desired precision of the results.

The statistical errors of the expansion coefficients $k_l$ of the
gradient-flow coupling, for example,
appear to grow rapidly with the loop order $l$.
In practice some loss of precision from one order to the next
is tolerable, since the coefficients get multiplied by $\alpha_s^{l+1}$
in the perturbation series (6.2).
An extension of the computations to three-loop order would however
only make sense if $k_1$ and $k_2$ are recalculated with errors about
an order of magnitude smaller than the ones quoted in table~3.
Furthermore, the relation between $\alpha_s$ and the
bare coupling would need to be worked out to three loops.

\vskip1.5ex

M.D.B.~would like to thank Marco Garofalo, Dirk Hesse, and Tony
Kennedy for a pleasant collaboration on related investigations,
Alberto Ramos, Stefan Sint, and Rainer Sommer for valuable discussions
and the Theoretical Physics Department at CERN for the kind hospitality
extended to him.
The computations reported in this paper were performed on
the SuperMUC machine at the Leibniz Supercomputing Centre
in Munich (project ID {\tt pr92ci}) and dedicated
HPC clusters at DESY-Zeuthen and at CERN.
We thank all these institutions for the generous support
given to this project.

\appendix A. OMF molecular-dynamics integrators

Among the popular integrators used in lattice QCD, there
are two schemes proposed by Omelyan, Mryglod and Folk (OMF) [\ref{OMF}],
one of second order in the integration step size and the other
of fourth order. Here their efficiency is studied by comparing
the distribution (4.8) actually simulated at leading order
of stochastic perturbation theory with the exact distribution.

\subsection A.1 2nd-order OMF integrator

The sequence of update steps (2.8),(2.9) is
\equation{
  I_{\eps}=I_{p,\eps_1}I_{q,\eps_2}I_{p,\eps_3}I_{q,\eps_2}I_{p,\eps_1},
  \quad \eps_k=r_k\eps,
  \enum
  \nexteq{2.5ex}
  r_1=0.1931833275037836,\quad r_2=1/2,\quad r_3=1-2r_1,
  \enum
}
in this case.
Using an algebraic manipulation program,
the expressions
\equation{
  M_{qp}=\eps\bigl(1-\frac{1}{4}r_3\eps^2\Op\bigr),
  \enum
  \nexteq{2.5ex}
  \Oph=\Op\bigl(1-\frac{1}{2}r_1\eps^2\Op\bigr)
          \bigl(1-\frac{1}{2}r_1r_3\eps^2\Op\bigr)
          \bigl(1-\frac{1}{4}r_3\eps^2\Op\bigr)^{-1},
  \enum
}
are then easily obtained.
The convergence of the SMD algorithm is thus guaranteed if
$\eps^2\norm{\Op}<6.51$.
With respect to the leapfrog integrator, and if the step size $\eps$
is adjusted so that the number of force evaluations per
unit time is the same, this scheme achieves
a reduction of the relative deviation of $\hat{\Op}$ from $\Op$ by a factor
$25$.

\subsection A.2 4th-order OMF integrator

This integrator is of the form (A.1) too, but with $11$ steps instead of
$5$. The associated step sizes in units of $\eps$ are
\equation{
  r_1=0.08398315262876693,
  \noenum
  \nexteq{1.0ex}
  r_2=0.2539785108410595,
  \noenum
  \nexteq{1.0ex}
  r_3=0.6822365335719091,
  \noenum
  \nexteq{1.0ex}
  r_4=-0.03230286765269967,
  \noenum
  \nexteq{1.0ex}
  r_5=1/2-r_1-r_3,\quad r_6=1-2(r_2+r_4),
  \enum
}
and the matrices $M_{pq}$ and $M_{qp}$ are polynomials in $\eps^2\Op$
of degree $5$ and $4$ times $\eps\Op$ and $\eps$, respectively.
In this case, $M_{qp}$ is non-singular and
\equation{
  \Oph=\Op\bigl\{1+a_1\eps^4\Op^2+a_2\eps^6\Op^3+\ldots\bigr\},
  \enum
  \nexteq{2.5ex}
  a_1=-2.58(1)\times10^{-5},\quad
  a_2=-1.88(1)\times10^{-5},
  \enum
}
is positive if $\eps^2\norm{\Op}<9.87$.

As is evident from eqs.~(A.6),(A.7),
this integrator is impressively accurate.
If $\norm{\Op}=16$ and $\eps=0.2$, for example,
the relative deviation of $\hat{\Op}$ from $\Op$ is at most $1.6\times10^{-5}$
and thus about a factor $12$ smaller than the
deviation in the case of the 2nd-order OMF integrator
(with an adjusted step size of $\eps=0.08$).

\appendix B. Explicit form of $\mib d^{\ast}$

The operator $d^{\ast}$ depends on the chosen
boundary conditions and the scalar products in the
field spaces $\Hs_0$ and $\Hs_1$ (cf.~subsect.~5.1).
For SF boundary conditions
\equation{
  (d^{\ast}\!A)(x)=-{T-x_0\over T^3L^3}
  \sum_{y_0=0}^{T-1}\sum_{\mib y}A_0(y)-
  \sum_{\mu=0}^3\drvstar{\mu}A_{\mu}(x),
  \enum
}
where
\equation{
  \drvstar{\mu}A_{\mu}(x)=\cases{
  A_{\mu}(x)-A_{\mu}(x-\hat{\mu}) & if $0<x_0<T$,\cr
  \noalign{\vskip1.0ex}
  0 & otherwise,\cr}
  \enum
}
while in the case of open-SF boundary conditions,
\equation{
  (d^{\ast}\!A)(x)=
  -\sum_{\mu=0}^3\drvstar{\mu}A_{\mu}(x),
  \enum
}
where now
\equation{
  \drvstar{\mu}A_{\mu}(x)=\cases{
  A_{\mu}(x)-A_{\mu}(x-\hat{\mu}) & if $\mu>0$ or $0<x_0<T$,\cr
  \noalign{\vskip1.0ex}
  2A_0(x) & if $\mu=0$ and $x_0=0$,\cr
  \noalign{\vskip1.0ex}
  0 & otherwise.\cr}
  \enum
}

\appendix C. Gauge-damped stochastic equations

In the continuous-time limit $\eps\to0$, the gauge-damped SMD algorithm
described in sect.~5 integrates the stochastic equations
\equation{
   \partial_tU_t(x,\mu)=g_0\left\{\pi_t(x,\mu)-\nab{\mu}\om_t(x)
   \right\}U_t(x,\mu),
   \enum
   \nexteq{2.5ex}
   \partial_t\pi_t(x,\mu)=-g_0w_{x,\mu}(\partial^a_{x,\mu}\SG)(U_t)T^a
   -\gamma\pi_t(x,\mu)+\eta_t(x,\mu)
   \noenum
   \nexteq{2.0ex}
   {\phantom{\partial_t\pi_t(x,\mu)={}}}
   +g_0\left[\om_t(x),\pi_t(x,\mu)\right],
   \enum
   \nexteq{2.5ex}
   \partial_t\om_t(x)=-\gamma\om_t(x)+\lambda_0(d^{\ast}\kern-1pt C_t)(x),
   \enum
}
where $\pi_t,U_t,\om_t$ denote
the momentum, gauge and gauge-damping fields at simulation time $t$
(and the link field $C_t$ is given by eq.~(5.23) with $U$ replaced by $U_t$).
The normalization of the Gaussian random field is such that
\equation{
   \langle\eta_t^a(x,\mu)\eta_s^b(y,\nu)\rangle=2\gamma
   w_{x,\mu}\delta^{ab}\delta_{\mu\nu}
   \delta(t-s)\delta_{xy}
   \enum
}
and
\equation{
  \nab{\mu}\om_t(x)=U_t(x,\mu)\om_t(x+\hat{\mu})U_t(x,\mu)^{-1}
  -\om_t(x)
  \enum
}
stands for the gauge-covariant gradient of the gauge-damping field.

As in the case of the SMD algorithm,
the gauge-damping terms can be removed
from the continuous-time process by applying a time-dependent gauge
transformation to the fields $\pi_t,U_t$ and $\eta_t$.
Up to a rescaling of the simulation time and the momentum field by
the coupling $g_0$, eqs.~(C.1),(C.2) then reduce to the standard stochastic
molecular-dynamics equations [\ref{Horowitz}].
The gauge damping can also easily be shown to coincide
with the standard one in the Langevin limit $\gamma\to\infty$
[\ref{Zwanziger}].

\vfill\eject

\appendix D. Tables of simulation results

{
\newdimen\digitwidth
\setbox0=\hbox{\rm 0}
\digitwidth=\wd0
\catcode`@=\active
\def@{\kern\digitwidth}
\tablecaption{Values of the expansion coefficient $k_1$ at $c=0.2$}
\vskip-3.0ex

$$\vbox{\settabs\+&%
                  xxx&xi&%
                  xxxxxxxxx&xx&%
                  xxxxxxxxx&xx&%
                  xxxxxxxxx&xx&%
                  xxxxxxxxx&xx&%
                  xxxxxxxxx&xx&%
                  xxxxxxxxx&xx\cr
\thicktablerule
\vskip1.2ex
                \+& \hfill $L$\hfill
                 && \hfill $k_1^{\rm w}$\hfill
                 && \hfill $k_1^{\rm c}$\hfill
                 && \hfill $k_1^{\rm ws}$\hfill
                 && \hfill $k_1^{\rm cs}$\hfill
                 && \hfill $k_1^{\rm wt}$\hfill
                 && \hfill $k_1^{\rm ct}$\hfill
                 &\cr
\vskip0.8ex
\thintablerule
\vskip1.2ex
{\ninepoint
  \+& \hfill $10$\hfill
  &&  \hfill $ 1.4876(9)@$\hfill
  &&  \hfill $ 1.3126(12)$\hfill
  &&  \hfill $ 1.5044(12)$\hfill
  &&  \hfill $ 1.3396(16)$\hfill
  &&  \hfill $ 1.4708(13)$\hfill
  &&  \hfill $ 1.2857(18)$\hfill
  &\cr
  \+& \hfill $12$\hfill
  &&  \hfill $ 1.3971(10)$\hfill
  &&  \hfill $ 1.2908(12)$\hfill
  &&  \hfill $ 1.4166(14)$\hfill
  &&  \hfill $ 1.3169(16)$\hfill
  &&  \hfill $ 1.3777(15)$\hfill
  &&  \hfill $ 1.2648(18)$\hfill
  &\cr
  \+& \hfill $14$\hfill
  &&  \hfill $ 1.3199(12)$\hfill
  &&  \hfill $ 1.2564(13)$\hfill
  &&  \hfill $ 1.3378(15)$\hfill
  &&  \hfill $ 1.2784(17)$\hfill
  &&  \hfill $ 1.3021(17)$\hfill
  &&  \hfill $ 1.2345(19)$\hfill
  &\cr
  \+& \hfill $16$\hfill
  &&  \hfill $ 1.2676(12)$\hfill
  &&  \hfill $ 1.2262(13)$\hfill
  &&  \hfill $ 1.2867(16)$\hfill
  &&  \hfill $ 1.2483(17)$\hfill
  &&  \hfill $ 1.2485(18)$\hfill
  &&  \hfill $ 1.2042(20)$\hfill
  &\cr
  \+& \hfill $18$\hfill
  &&  \hfill $ 1.2319(13)$\hfill
  &&  \hfill $ 1.2026(14)$\hfill
  &&  \hfill $ 1.2518(17)$\hfill
  &&  \hfill $ 1.2249(18)$\hfill
  &&  \hfill $ 1.2121(19)$\hfill
  &&  \hfill $ 1.1804(21)$\hfill
  &\cr
  \+& \hfill $20$\hfill
  &&  \hfill $ 1.2063(13)$\hfill
  &&  \hfill $ 1.1842(14)$\hfill
  &&  \hfill $ 1.2270(18)$\hfill
  &&  \hfill $ 1.2069(18)$\hfill
  &&  \hfill $ 1.1857(20)$\hfill
  &&  \hfill $ 1.1617(21)$\hfill
  &\cr
  \+& \hfill $24$\hfill
  &&  \hfill $ 1.1725(14)$\hfill
  &&  \hfill $ 1.1587(14)$\hfill
  &&  \hfill $ 1.1910(18)$\hfill
  &&  \hfill $ 1.1784(18)$\hfill
  &&  \hfill $ 1.1540(20)$\hfill
  &&  \hfill $ 1.1390(21)$\hfill
  &\cr
  \+& \hfill $28$\hfill
  &&  \hfill $ 1.1561(14)$\hfill
  &&  \hfill $ 1.1466(14)$\hfill
  &&  \hfill $ 1.1756(18)$\hfill
  &&  \hfill $ 1.1670(18)$\hfill
  &&  \hfill $ 1.1367(21)$\hfill
  &&  \hfill $ 1.1263(21)$\hfill
  &\cr
  \+& \hfill $32$\hfill
  &&  \hfill $ 1.1369(14)$\hfill
  &&  \hfill $ 1.1298(14)$\hfill
  &&  \hfill $ 1.1578(18)$\hfill
  &&  \hfill $ 1.1514(18)$\hfill
  &&  \hfill $ 1.1161(20)$\hfill
  &&  \hfill $ 1.1083(21)$\hfill
  &\cr
  \+& \hfill $40$\hfill
  &&  \hfill $ 1.1202(15)$\hfill
  &&  \hfill $ 1.1159(15)$\hfill
  &&  \hfill $ 1.1401(19)$\hfill
  &&  \hfill $ 1.1363(19)$\hfill
  &&  \hfill $ 1.1004(22)$\hfill
  &&  \hfill $ 1.0957(22)$\hfill
  &\cr
}
\vskip1.2ex
\thicktablerule
}
$$
\vskip0.0ex
}

\vskip0.5cm

{
\newdimen\digitwidth
\setbox0=\hbox{\rm 0}
\digitwidth=\wd0
\catcode`@=\active
\def@{\kern\digitwidth}
\tablecaption{Values of the expansion coefficient $k_1$ at $c=0.3$}
\vskip-3.0ex

$$\vbox{\settabs\+&%
                  xxx&xi&%
                  xxxxxxxxx&xx&%
                  xxxxxxxxx&xx&%
                  xxxxxxxxx&xx&%
                  xxxxxxxxx&xx&%
                  xxxxxxxxx&xx&%
                  xxxxxxxxx&xx\cr
\thicktablerule
\vskip1.2ex
                \+& \hfill $L$\hfill
                 && \hfill $k_1^{\rm w}$\hfill
                 && \hfill $k_1^{\rm c}$\hfill
                 && \hfill $k_1^{\rm ws}$\hfill
                 && \hfill $k_1^{\rm cs}$\hfill
                 && \hfill $k_1^{\rm wt}$\hfill
                 && \hfill $k_1^{\rm ct}$\hfill
                 &\cr
\vskip0.8ex
\thintablerule
\vskip1.2ex
{\ninepoint
  \+& \hfill $10$\hfill
  &&  \hfill $ 1.3120(21)$\hfill
  &&  \hfill $ 1.3062(24)$\hfill
  &&  \hfill $ 1.4241(28)$\hfill
  &&  \hfill $ 1.4409(30)$\hfill
  &&  \hfill $ 1.2021(30)$\hfill
  &&  \hfill $ 1.1745(34)$\hfill
  &\cr
  \+& \hfill $12$\hfill
  &&  \hfill $ 1.2539(23)$\hfill
  &&  \hfill $ 1.2545(24)$\hfill
  &&  \hfill $ 1.3710(31)$\hfill
  &&  \hfill $ 1.3869(32)$\hfill
  &&  \hfill $ 1.1391(32)$\hfill
  &&  \hfill $ 1.1250(35)$\hfill
  &\cr
  \+& \hfill $14$\hfill
  &&  \hfill $ 1.2200(25)$\hfill
  &&  \hfill $ 1.2225(26)$\hfill
  &&  \hfill $ 1.3316(32)$\hfill
  &&  \hfill $ 1.3448(33)$\hfill
  &&  \hfill $ 1.1105(36)$\hfill
  &&  \hfill $ 1.1027(38)$\hfill
  &\cr
  \+& \hfill $16$\hfill
  &&  \hfill $ 1.1952(26)$\hfill
  &&  \hfill $ 1.1977(27)$\hfill
  &&  \hfill $ 1.3061(33)$\hfill
  &&  \hfill $ 1.3164(34)$\hfill
  &&  \hfill $ 1.0866(38)$\hfill
  &&  \hfill $ 1.0816(39)$\hfill
  &\cr
  \+& \hfill $18$\hfill
  &&  \hfill $ 1.1800(27)$\hfill
  &&  \hfill $ 1.1823(28)$\hfill
  &&  \hfill $ 1.2937(35)$\hfill
  &&  \hfill $ 1.3023(36)$\hfill
  &&  \hfill $ 1.0685(39)$\hfill
  &&  \hfill $ 1.0647(40)$\hfill
  &\cr
  \+& \hfill $20$\hfill
  &&  \hfill $ 1.1671(27)$\hfill
  &&  \hfill $ 1.1691(28)$\hfill
  &&  \hfill $ 1.2799(35)$\hfill
  &&  \hfill $ 1.2870(36)$\hfill
  &&  \hfill $ 1.0566(40)$\hfill
  &&  \hfill $ 1.0537(41)$\hfill
  &\cr
  \+& \hfill $24$\hfill
  &&  \hfill $ 1.1515(27)$\hfill
  &&  \hfill $ 1.1532(27)$\hfill
  &&  \hfill $ 1.2595(35)$\hfill
  &&  \hfill $ 1.2646(35)$\hfill
  &&  \hfill $ 1.0456(39)$\hfill
  &&  \hfill $ 1.0441(40)$\hfill
  &\cr
  \+& \hfill $28$\hfill
  &&  \hfill $ 1.1468(28)$\hfill
  &&  \hfill $ 1.1482(28)$\hfill
  &&  \hfill $ 1.2593(36)$\hfill
  &&  \hfill $ 1.2632(36)$\hfill
  &&  \hfill $ 1.0365(41)$\hfill
  &&  \hfill $ 1.0354(42)$\hfill
  &\cr
  \+& \hfill $32$\hfill
  &&  \hfill $ 1.1312(28)$\hfill
  &&  \hfill $ 1.1322(28)$\hfill
  &&  \hfill $ 1.2444(35)$\hfill
  &&  \hfill $ 1.2474(36)$\hfill
  &&  \hfill $ 1.0203(41)$\hfill
  &&  \hfill $ 1.0194(41)$\hfill
  &\cr
  \+& \hfill $40$\hfill
  &&  \hfill $ 1.1264(30)$\hfill
  &&  \hfill $ 1.1271(30)$\hfill
  &&  \hfill $ 1.2359(39)$\hfill
  &&  \hfill $ 1.2378(39)$\hfill
  &&  \hfill $ 1.0192(45)$\hfill
  &&  \hfill $ 1.0187(45)$\hfill
  &\cr
}
\vskip1.2ex
\thicktablerule
}
$$
\vskip0.0ex
}

\vfill\eject

{
\newdimen\digitwidth
\setbox0=\hbox{\rm 0}
\digitwidth=\wd0
\catcode`@=\active
\def@{\kern\digitwidth}
\tablecaption{Values of the expansion coefficient $k_1$ at $c=0.4$}
\vskip-3.0ex

$$\vbox{\settabs\+&%
                  xxx&xi&%
                  xxxxxxxxx&xx&%
                  xxxxxxxxx&xx&%
                  xxxxxxxxx&xx&%
                  xxxxxxxxx&xx&%
                  xxxxxxxxx&xx&%
                  xxxxxxxxx&xx\cr
\thicktablerule
\vskip1.2ex
                \+& \hfill $L$\hfill
                 && \hfill $k_1^{\rm w}$\hfill
                 && \hfill $k_1^{\rm c}$\hfill
                 && \hfill $k_1^{\rm ws}$\hfill
                 && \hfill $k_1^{\rm cs}$\hfill
                 && \hfill $k_1^{\rm wt}$\hfill
                 && \hfill $k_1^{\rm ct}$\hfill
                 &\cr
\vskip0.8ex
\thintablerule
\vskip1.2ex
{\ninepoint
  \+& \hfill $10$\hfill
  &&  \hfill $ 1.3582(35)$\hfill
  &&  \hfill $ 1.4408(38)$\hfill
  &&  \hfill $ 1.7545(48)$\hfill
  &&  \hfill $ 1.8741(50)$\hfill
  &&  \hfill $ 0.9928(49)$\hfill
  &&  \hfill $ 1.0415(53)$\hfill
  &\cr
  \+& \hfill $12$\hfill
  &&  \hfill $ 1.3373(37)$\hfill
  &&  \hfill $ 1.3953(39)$\hfill
  &&  \hfill $ 1.7405(52)$\hfill
  &&  \hfill $ 1.8244(53)$\hfill
  &&  \hfill $ 0.9651(52)$\hfill
  &&  \hfill $ 0.9994(55)$\hfill
  &\cr
  \+& \hfill $14$\hfill
  &&  \hfill $ 1.3337(41)$\hfill
  &&  \hfill $ 1.3772(43)$\hfill
  &&  \hfill $ 1.7303(55)$\hfill
  &&  \hfill $ 1.7927(57)$\hfill
  &&  \hfill $ 0.9667(57)$\hfill
  &&  \hfill $ 0.9930(59)$\hfill
  &\cr
  \+& \hfill $16$\hfill
  &&  \hfill $ 1.3197(43)$\hfill
  &&  \hfill $ 1.3527(44)$\hfill
  &&  \hfill $ 1.7097(57)$\hfill
  &&  \hfill $ 1.7568(58)$\hfill
  &&  \hfill $ 0.9592(61)$\hfill
  &&  \hfill $ 0.9793(63)$\hfill
  &\cr
  \+& \hfill $18$\hfill
  &&  \hfill $ 1.3183(44)$\hfill
  &&  \hfill $ 1.3444(45)$\hfill
  &&  \hfill $ 1.7137(59)$\hfill
  &&  \hfill $ 1.7513(60)$\hfill
  &&  \hfill $ 0.9525(61)$\hfill
  &&  \hfill $ 0.9681(63)$\hfill
  &\cr
  \+& \hfill $20$\hfill
  &&  \hfill $ 1.3181(45)$\hfill
  &&  \hfill $ 1.3396(46)$\hfill
  &&  \hfill $ 1.7111(59)$\hfill
  &&  \hfill $ 1.7419(59)$\hfill
  &&  \hfill $ 0.9543(64)$\hfill
  &&  \hfill $ 0.9673(65)$\hfill
  &\cr
  \+& \hfill $24$\hfill
  &&  \hfill $ 1.3121(44)$\hfill
  &&  \hfill $ 1.3271(44)$\hfill
  &&  \hfill $ 1.6983(58)$\hfill
  &&  \hfill $ 1.7197(58)$\hfill
  &&  \hfill $ 0.9544(61)$\hfill
  &&  \hfill $ 0.9636(62)$\hfill
  &\cr
  \+& \hfill $28$\hfill
  &&  \hfill $ 1.3116(46)$\hfill
  &&  \hfill $ 1.3227(47)$\hfill
  &&  \hfill $ 1.7068(59)$\hfill
  &&  \hfill $ 1.7226(60)$\hfill
  &&  \hfill $ 0.9456(66)$\hfill
  &&  \hfill $ 0.9523(67)$\hfill
  &\cr
  \+& \hfill $32$\hfill
  &&  \hfill $ 1.2987(46)$\hfill
  &&  \hfill $ 1.3070(47)$\hfill
  &&  \hfill $ 1.6918(60)$\hfill
  &&  \hfill $ 1.7038(60)$\hfill
  &&  \hfill $ 0.9349(65)$\hfill
  &&  \hfill $ 0.9399(66)$\hfill
  &\cr
  \+& \hfill $40$\hfill
  &&  \hfill $ 1.3012(50)$\hfill
  &&  \hfill $ 1.3066(51)$\hfill
  &&  \hfill $ 1.6876(65)$\hfill
  &&  \hfill $ 1.6953(65)$\hfill
  &&  \hfill $ 0.9433(72)$\hfill
  &&  \hfill $ 0.9467(72)$\hfill
  &\cr
}
\vskip1.2ex
\thicktablerule
}
$$
\vskip0.0ex
}

\vskip0.5cm

{
\newdimen\digitwidth
\setbox0=\hbox{\rm 0}
\digitwidth=\wd0
\catcode`@=\active
\def@{\kern\digitwidth}
\tablecaption{Values of the expansion coefficient $k_2$ at $c=0.2$}
\vskip-3.0ex

$$\vbox{\settabs\+&%
                  xxx&xi&%
                  xxxxxxxxx&xx&%
                  xxxxxxxxx&xx&%
                  xxxxxxxxx&xx&%
                  xxxxxxxxx&xx&%
                  xxxxxxxxx&xx&%
                  xxxxxxxxx&xx\cr
\thicktablerule
\vskip1.2ex
                \+& \hfill $L$\hfill
                 && \hfill $k_2^{\rm w}$\hfill
                 && \hfill $k_2^{\rm c}$\hfill
                 && \hfill $k_2^{\rm ws}$\hfill
                 && \hfill $k_2^{\rm cs}$\hfill
                 && \hfill $k_2^{\rm wt}$\hfill
                 && \hfill $k_2^{\rm ct}$\hfill
                 &\cr
\vskip0.8ex
\thintablerule
\vskip1.2ex
{\ninepoint
  \+& \hfill $10$\hfill
  &&  \hfill $-1.863(9)@$\hfill
  &&  \hfill $-1.893(12)$\hfill
  &&  \hfill $-1.919(12)$\hfill
  &&  \hfill $-1.985(16)$\hfill
  &&  \hfill $-1.807(12)$\hfill
  &&  \hfill $-1.801(18)$\hfill
  &\cr
  \+& \hfill $12$\hfill
  &&  \hfill $-1.760(11)$\hfill
  &&  \hfill $-1.742(13)$\hfill
  &&  \hfill $-1.820(14)$\hfill
  &&  \hfill $-1.826(17)$\hfill
  &&  \hfill $-1.700(15)$\hfill
  &&  \hfill $-1.658(19)$\hfill
  &\cr
  \+& \hfill $14$\hfill
  &&  \hfill $-1.706(12)$\hfill
  &&  \hfill $-1.679(14)$\hfill
  &&  \hfill $-1.755(16)$\hfill
  &&  \hfill $-1.739(17)$\hfill
  &&  \hfill $-1.657(17)$\hfill
  &&  \hfill $-1.620(20)$\hfill
  &\cr
  \+& \hfill $16$\hfill
  &&  \hfill $-1.655(13)$\hfill
  &&  \hfill $-1.630(14)$\hfill
  &&  \hfill $-1.718(17)$\hfill
  &&  \hfill $-1.702(18)$\hfill
  &&  \hfill $-1.593(19)$\hfill
  &&  \hfill $-1.558(21)$\hfill
  &\cr
  \+& \hfill $18$\hfill
  &&  \hfill $-1.608(15)$\hfill
  &&  \hfill $-1.586(16)$\hfill
  &&  \hfill $-1.674(19)$\hfill
  &&  \hfill $-1.660(20)$\hfill
  &&  \hfill $-1.542(22)$\hfill
  &&  \hfill $-1.512(24)$\hfill
  &\cr
  \+& \hfill $20$\hfill
  &&  \hfill $-1.562(16)$\hfill
  &&  \hfill $-1.544(17)$\hfill
  &&  \hfill $-1.611(20)$\hfill
  &&  \hfill $-1.597(21)$\hfill
  &&  \hfill $-1.513(24)$\hfill
  &&  \hfill $-1.490(25)$\hfill
  &\cr
  \+& \hfill $24$\hfill
  &&  \hfill $-1.486(17)$\hfill
  &&  \hfill $-1.474(18)$\hfill
  &&  \hfill $-1.534(22)$\hfill
  &&  \hfill $-1.526(23)$\hfill
  &&  \hfill $-1.438(25)$\hfill
  &&  \hfill $-1.423(27)$\hfill
  &\cr
  \+& \hfill $28$\hfill
  &&  \hfill $-1.473(18)$\hfill
  &&  \hfill $-1.465(19)$\hfill
  &&  \hfill $-1.520(24)$\hfill
  &&  \hfill $-1.515(24)$\hfill
  &&  \hfill $-1.426(28)$\hfill
  &&  \hfill $-1.415(29)$\hfill
  &\cr
  \+& \hfill $32$\hfill
  &&  \hfill $-1.365(18)$\hfill
  &&  \hfill $-1.357(19)$\hfill
  &&  \hfill $-1.438(24)$\hfill
  &&  \hfill $-1.433(24)$\hfill
  &&  \hfill $-1.293(28)$\hfill
  &&  \hfill $-1.282(28)$\hfill
  &\cr
  \+& \hfill $40$\hfill
  &&  \hfill $-1.317(21)$\hfill
  &&  \hfill $-1.312(21)$\hfill
  &&  \hfill $-1.398(28)$\hfill
  &&  \hfill $-1.395(28)$\hfill
  &&  \hfill $-1.236(32)$\hfill
  &&  \hfill $-1.230(32)$\hfill
  &\cr
}
\vskip1.2ex
\thicktablerule
}
$$
\vskip0.0ex
}

\vfill\eject

{
\newdimen\digitwidth
\setbox0=\hbox{\rm 0}
\digitwidth=\wd0
\catcode`@=\active
\def@{\kern\digitwidth}
\tablecaption{Values of the expansion coefficient $k_2$ at $c=0.3$}
\vskip-3.0ex

$$\vbox{\settabs\+&%
                  xxx&xi&%
                  xxxxxxxxx&xx&%
                  xxxxxxxxx&xx&%
                  xxxxxxxxx&xx&%
                  xxxxxxxxx&xx&%
                  xxxxxxxxx&xx&%
                  xxxxxxxxx&xx\cr
\thicktablerule
\vskip1.2ex
                \+& \hfill $L$\hfill
                 && \hfill $k_2^{\rm w}$\hfill
                 && \hfill $k_2^{\rm c}$\hfill
                 && \hfill $k_2^{\rm ws}$\hfill
                 && \hfill $k_2^{\rm cs}$\hfill
                 && \hfill $k_2^{\rm wt}$\hfill
                 && \hfill $k_2^{\rm ct}$\hfill
                 &\cr
\vskip0.8ex
\thintablerule
\vskip1.2ex
{\ninepoint
  \+& \hfill $10$\hfill
  &&  \hfill $-2.134(23)$\hfill
  &&  \hfill $-2.229(26)$\hfill
  &&  \hfill $-2.582(31)$\hfill
  &&  \hfill $-2.765(33)$\hfill
  &&  \hfill $-1.695(33)$\hfill
  &&  \hfill $-1.706(37)$\hfill
  &\cr
  \+& \hfill $12$\hfill
  &&  \hfill $-2.036(25)$\hfill
  &&  \hfill $-2.097(27)$\hfill
  &&  \hfill $-2.502(34)$\hfill
  &&  \hfill $-2.618(36)$\hfill
  &&  \hfill $-1.580(37)$\hfill
  &&  \hfill $-1.587(40)$\hfill
  &\cr
  \+& \hfill $14$\hfill
  &&  \hfill $-2.036(28)$\hfill
  &&  \hfill $-2.084(30)$\hfill
  &&  \hfill $-2.469(36)$\hfill
  &&  \hfill $-2.555(38)$\hfill
  &&  \hfill $-1.611(41)$\hfill
  &&  \hfill $-1.624(44)$\hfill
  &\cr
  \+& \hfill $16$\hfill
  &&  \hfill $-1.999(30)$\hfill
  &&  \hfill $-2.034(31)$\hfill
  &&  \hfill $-2.427(38)$\hfill
  &&  \hfill $-2.488(40)$\hfill
  &&  \hfill $-1.580(45)$\hfill
  &&  \hfill $-1.589(47)$\hfill
  &\cr
  \+& \hfill $18$\hfill
  &&  \hfill $-1.944(33)$\hfill
  &&  \hfill $-1.969(34)$\hfill
  &&  \hfill $-2.385(42)$\hfill
  &&  \hfill $-2.433(43)$\hfill
  &&  \hfill $-1.511(49)$\hfill
  &&  \hfill $-1.514(50)$\hfill
  &\cr
  \+& \hfill $20$\hfill
  &&  \hfill $-1.919(35)$\hfill
  &&  \hfill $-1.939(36)$\hfill
  &&  \hfill $-2.349(44)$\hfill
  &&  \hfill $-2.388(45)$\hfill
  &&  \hfill $-1.498(52)$\hfill
  &&  \hfill $-1.500(54)$\hfill
  &\cr
  \+& \hfill $24$\hfill
  &&  \hfill $-1.889(36)$\hfill
  &&  \hfill $-1.905(37)$\hfill
  &&  \hfill $-2.267(48)$\hfill
  &&  \hfill $-2.294(48)$\hfill
  &&  \hfill $-1.518(54)$\hfill
  &&  \hfill $-1.524(55)$\hfill
  &\cr
  \+& \hfill $28$\hfill
  &&  \hfill $-1.912(40)$\hfill
  &&  \hfill $-1.924(40)$\hfill
  &&  \hfill $-2.337(51)$\hfill
  &&  \hfill $-2.357(51)$\hfill
  &&  \hfill $-1.497(60)$\hfill
  &&  \hfill $-1.500(60)$\hfill
  &\cr
  \+& \hfill $32$\hfill
  &&  \hfill $-1.770(40)$\hfill
  &&  \hfill $-1.778(40)$\hfill
  &&  \hfill $-2.222(51)$\hfill
  &&  \hfill $-2.238(51)$\hfill
  &&  \hfill $-1.327(60)$\hfill
  &&  \hfill $-1.328(61)$\hfill
  &\cr
  \+& \hfill $40$\hfill
  &&  \hfill $-1.794(45)$\hfill
  &&  \hfill $-1.800(46)$\hfill
  &&  \hfill $-2.229(58)$\hfill
  &&  \hfill $-2.239(58)$\hfill
  &&  \hfill $-1.368(68)$\hfill
  &&  \hfill $-1.370(69)$\hfill
  &\cr
}
\vskip1.2ex
\thicktablerule
}
$$
\vskip0.0ex
}

\vskip0.5cm

{
\newdimen\digitwidth
\setbox0=\hbox{\rm 0}
\digitwidth=\wd0
\catcode`@=\active
\def@{\kern\digitwidth}
\tablecaption{Values of the expansion coefficient $k_2$ at $c=0.4$}
\vskip-3.0ex

$$\vbox{\settabs\+&%
                  xxx&xi&%
                  xxxxxxxxx&xx&%
                  xxxxxxxxx&xx&%
                  xxxxxxxxx&xx&%
                  xxxxxxxxx&xx&%
                  xxxxxxxxx&xx&%
                  xxxxxxxxx&xx\cr
\thicktablerule
\vskip1.2ex
                \+& \hfill $L$\hfill
                 && \hfill $k_2^{\rm w}$\hfill
                 && \hfill $k_2^{\rm c}$\hfill
                 && \hfill $k_2^{\rm ws}$\hfill
                 && \hfill $k_2^{\rm cs}$\hfill
                 && \hfill $k_2^{\rm wt}$\hfill
                 && \hfill $k_2^{\rm ct}$\hfill
                 &\cr
\vskip0.8ex
\thintablerule
\vskip1.2ex
{\ninepoint
  \+& \hfill $10$\hfill
  &&  \hfill $-3.118(41)$\hfill
  &&  \hfill $-3.417(45)$\hfill
  &&  \hfill $-4.908(56)$\hfill
  &&  \hfill $-5.384(59)$\hfill
  &&  \hfill $-1.468(58)$\hfill
  &&  \hfill $-1.605(63)$\hfill
  &\cr
  \+& \hfill $12$\hfill
  &&  \hfill $-3.039(45)$\hfill
  &&  \hfill $-3.248(47)$\hfill
  &&  \hfill $-4.858(61)$\hfill
  &&  \hfill $-5.183(63)$\hfill
  &&  \hfill $-1.361(64)$\hfill
  &&  \hfill $-1.463(67)$\hfill
  &\cr
  \+& \hfill $14$\hfill
  &&  \hfill $-3.164(50)$\hfill
  &&  \hfill $-3.321(52)$\hfill
  &&  \hfill $-4.950(67)$\hfill
  &&  \hfill $-5.189(70)$\hfill
  &&  \hfill $-1.511(70)$\hfill
  &&  \hfill $-1.594(73)$\hfill
  &\cr
  \+& \hfill $16$\hfill
  &&  \hfill $-3.124(53)$\hfill
  &&  \hfill $-3.241(55)$\hfill
  &&  \hfill $-4.863(71)$\hfill
  &&  \hfill $-5.041(73)$\hfill
  &&  \hfill $-1.52(8)@@$\hfill
  &&  \hfill $-1.58(8)@@$\hfill
  &\cr
  \+& \hfill $18$\hfill
  &&  \hfill $-3.054(58)$\hfill
  &&  \hfill $-3.142(59)$\hfill
  &&  \hfill $-4.82(8)@@$\hfill
  &&  \hfill $-4.96(8)@@$\hfill
  &&  \hfill $-1.42(8)@@$\hfill
  &&  \hfill $-1.46(8)@@$\hfill
  &\cr
  \+& \hfill $20$\hfill
  &&  \hfill $-3.106(61)$\hfill
  &&  \hfill $-3.179(63)$\hfill
  &&  \hfill $-4.88(8)@@$\hfill
  &&  \hfill $-4.99(8)@@$\hfill
  &&  \hfill $-1.47(9)@@$\hfill
  &&  \hfill $-1.50(9)@@$\hfill
  &\cr
  \+& \hfill $24$\hfill
  &&  \hfill $-3.080(62)$\hfill
  &&  \hfill $-3.131(63)$\hfill
  &&  \hfill $-4.75(8)@@$\hfill
  &&  \hfill $-4.83(8)@@$\hfill
  &&  \hfill $-1.54(9)@@$\hfill
  &&  \hfill $-1.56(9)@@$\hfill
  &\cr
  \+& \hfill $28$\hfill
  &&  \hfill $-3.125(69)$\hfill
  &&  \hfill $-3.163(70)$\hfill
  &&  \hfill $-4.89(9)@@$\hfill
  &&  \hfill $-4.95(9)@@$\hfill
  &&  \hfill $-1.49(10)@$\hfill
  &&  \hfill $-1.51(10)@$\hfill
  &\cr
  \+& \hfill $32$\hfill
  &&  \hfill $-2.996(70)$\hfill
  &&  \hfill $-3.025(71)$\hfill
  &&  \hfill $-4.79(9)@@$\hfill
  &&  \hfill $-4.84(9)@@$\hfill
  &&  \hfill $-1.33(10)@$\hfill
  &&  \hfill $-1.35(10)@$\hfill
  &\cr
  \+& \hfill $40$\hfill
  &&  \hfill $-3.06(8)@@$\hfill
  &&  \hfill $-3.08(8)@@$\hfill
  &&  \hfill $-4.77(10)@$\hfill
  &&  \hfill $-4.79(10)@$\hfill
  &&  \hfill $-1.49(11)@$\hfill
  &&  \hfill $-1.50(12)@$\hfill
  &\cr
}
\vskip1.2ex
\thicktablerule
}
$$
\vskip0.0ex
}

\vfill\eject

\beginbibliography


\bibitem{SPThI}
F. Di Renzo, G. Marchesini, P. Marenzoni, E. Onofri,
{\it Lattice perturbation theory on the computer},
Nucl. Phys. Proc. Suppl. 34, 795 (1994).

\bibitem{SPThII}
F. Di Renzo, E. Onofri, G. Marchesini, P. Marenzoni,
{\it Four loop result in SU(3) lattice gauge theory by a stochastic method:
Lattice correction to the condensate},
Nucl. Phys. B426, 675 (1994).


\bibitem{DiRenzoScorzato}
F. Di Renzo, L. Scorzato,
{\it Numerical stochastic perturbation theory for full QCD},
JHEP 0410 (2004) 073.


\bibitem{HMC}
S. Duane, A. D. Kennedy, B. J. Pendleton, D. Roweth,
{\it Hybrid Monte Carlo},
Phys. Lett. B195 (1987) 216.


\bibitem{Horowitz}
A. M. Horowitz,
{\it Stochastic quantization in phase space},
Phys. Lett. 156B (1985) 89;
{\it The second order Langevin equation and numerical simulations},
Nucl. Phys. B280 [FS18] (1987) 510;
{\it A generalized guided Monte Carlo algorithm},
Phys. Lett. B268 (1991) 247.


\bibitem{LatKobe}
M. Dalla Brida, M. Garofalo, A. D. Kennedy,
{\it
Numerical stochastic perturbation theory and gradient flow
in $\phi^4$ theory},
PoS LATTICE2015 (2016) 309.


\bibitem{JansenLiu}
K. Jansen, C. Liu,
{\it Kramers equation algorithm for simulations of QCD
with two flavors of Wilson fermions and gauge group SU(2)},
Nucl. Phys. B453 (1995) 375 [Erratum: {\it ibid.} B459 (1996) 437].


\bibitem{OMF}
I. P. Omelyan, I. M. Mryglod, R. Folk,
{\it Symplectic analytically integrable decomposition
algorithms: classification, derivation, and application to molecular
dynamics, quantum and celestial mechanics simulations},
Comp. Phys. Commun. 151 (2003) 272.


\bibitem{SQI}
G. Parisi, Y.-S. Wu,
{\it Perturbation theory without gauge fixing},
Sci. Sin. 24 (1981) 483.

\bibitem{SQII}
P. H. Damgaard, H. H\"uffel,
{\it Stochastic quantization},
Phys. Rept. 152 (1987) 227.


\bibitem{SF}
M. L\"uscher, R. Narayanan, P. Weisz, U. Wolff,
{\it The Schr\"odinger functional ---
a renormalizable probe for non-Abelian gauge theories},
Nucl. Phys. B384 (1992) 168.

\bibitem{SFquark}
S. Sint,
{\it On the Schr\"odinger functional in QCD},
Nucl. Phys. B421 (1994) 135.


\bibitem{openSF}
M. L\"uscher,
{\it Step scaling and the Yang--Mills gradient flow},
JHEP 1406 (2014) 105.


\bibitem{Wilson}
K. G. Wilson, {\it Confinement of quarks}, Phys. Rev. D10 (1974) 2445.

\bibitem{SymImpI}
P. Weisz,
{\it Continuum limit improved lattice action for pure Yang--Mills theory (I)},
Nucl. Phys. B212 (1983) 1.

\bibitem{OnShell}
M. L\"uscher, P. Weisz,
{\it On-shell improved lattice gauge theories},
Commun. Math. Phys. 97 (1985) 59 [Erratum: {\it ibid.} 98 (1985) 433].

\bibitem{Iwasaki}
Y. Iwasaki,
{\it Renormalization group analysis of lattice theories and
improved lattice action. II -- Four-dimensional non-Abelian SU(N) gauge model},
preprint UTHEP-118 (1983) [arXiv:1111.7054v1].


\bibitem{LesHouches}
M.~L\"uscher,
{\it Computational strategies in lattice QCD},
in: {\it Modern perspectives in lattice QCD},
eds. L. Lellouch et al. (Oxford University Press, New York, 2011).


\bibitem{FritzschRamos}
P. Fritzsch, A. Ramos,
{\it The gradient flow coupling in the Schr\"odinger functional},
JHEP 1310 (2013) 008.


\bibitem{StepQCDI}
M. Dalla Brida et al. (ALPHA Collab.),
{\it Slow running of the gradient flow coupling from 200 MeV to 4 GeV
in $N_f=3$ QCD},
Phys. Rev. D95 (2017) 014507.

\bibitem{StepQCDII}
I. Campos et al.,
{\it Non-perturbative running of quark masses in three-flavour QCD},
PoS LATTICE2016 (2016) 201.


\bibitem{StepScaling}
M. L\"uscher, P. Weisz, U. Wolff,
{\it A numerical method to compute the running coupling in asymptotically
free theories},
Nucl. Phys. B359 (1991) 221.

\bibitem{HouchesII}
M. L\"uscher,
{\it Advanced lattice QCD},
in: Probing the
Standard Model of Particle Interactions (Les Houches 1997),
eds. R. Gupta et al.
(Elsevier, Amsterdam, 1999).


\bibitem{WilsonFlow}
M. L\"uscher,
{\it Properties and uses of the Wilson flow in lattice QCD},
JHEP 1008 (2010) 071 [Erratum: {\it ibid.} 1403 (2014) 092].


\bibitem{HarlanderNeumann}
R. V. Harlander, T. Neumann,
{\it The perturbative QCD gradient flow to three loops},
JHEP 1606 (2016) 161.


\bibitem{RenFlow}
M. L\"uscher, P. Weisz,
{\it Perturbative analysis of the gradient flow in non-Abelian gauge
theories},
JHEP 1102 (2011) 051.


\bibitem{LatMSbar}
M. L\"uscher, P. Weisz,
{\it Two-loop relation between the bare lattice coupling and the\/
$\MSbar$ coupling in pure $\SUn$ gauge theories},
Phys. Lett. B349 (1995) 165.


\bibitem{CtI}
M. L\"uscher, R. Sommer, P. Weisz, U. Wolff,
{\it A precise determination of the running coupling in the
SU(3) Yang--Mills theory},
Nucl. Phys. B413 (1994) 481.

\bibitem{CtII}
A. Bode, U. Wolff, P. Weisz,
{\it Two-loop computation of the Schr\"odinger
functional in pure SU(3) lattice gauge theory},
Nucl.Phys. B540 (1999) 491.

\bibitem{CtIII}
A. Bode, P. Weisz, U. Wolff,
{\it Two-loop computation of the Schr\"odinger
functional in lattice QCD},
Nucl. Phys. B576 (2000) 517
[Errata: {\it ibid.} B600 (2001) 453,
{\it ibid.} B608 (2001) 481].


\bibitem{PhiFour}
M. Dalla Brida, M. Garofalo, A. D. Kennedy,
{\it An investigation of new methods for numerical stochastic perturbation
theory in $\varphi^4$ theory}, arXiv:1703.04406 [hep-lat]


\bibitem{Zinn}
J. Zinn--Justin,
{\it Renormalization and stochastic quantization},
Nucl. Phys. B275 [FS17] (1986) 135.

\bibitem{ZinnZwanziger}
J. Zinn--Justin, D. Zwanziger,
{\it Ward identities for the stochastic quantization of gauge fields},
Nucl. Phys. B295 [FS21] (1988) 297.

\bibitem{LangNotes}
M. L\"uscher,
{\it Statistical errors in stochastic perturbation theory},
notes (2015),
{\tt http://luscher.web.cern.ch/luscher/notes/enspt.pdf}.


\bibitem{SymI}
K. Symanzik,
{\it Cutoff dependence in lattice $\phi^4_4$ theory},
in: Recent developments in gauge theories (Carg\`ese 1979),
eds. G. 't Hooft et al. (Plenum, New York, 1980).

\bibitem{SymII}
K. Symanzik,
{\it Some topics in quantum field theory},
in: Mathematical problems in theoretical physics,
eds. R. Schrader et al., Lecture Notes in Physics, Vol. 153
(Springer, New York, 1982).


\bibitem{Zwanziger}
D. Zwanziger, {\it Covariant quantization of gauge fields
without Gribov ambiguity},
Nucl. Phys. B192 (1981) 259.

\endbibliography

\bye